# Spherical Radiomics – A Novel Approach to Glioblastoma Radiogenomic Analysis of Heterogeneity

Haotian Feng,[1] and Ke Sheng[1,*]

[1]Department of Radiation Oncology, University of California, San Francisco. San Francisco, California, USA
*Correspondence: Ke.Sheng@ucsf.edu

## SUMMARY

**Purpose:** To develop and validate a novel spherical radiomics framework for predicting key molecular biomarkers—including MGMT promoter methylation, EGFR, and PTEN mutation status—as well as survival for glioblastoma (GBM) patients using multiparametric MRI.

**Methods:** Conventional Cartesian radiomics extract tumor features on orthogonal grids, which do not fully capture the tumor's radial growth patterns and can be insensitive to evolving molecular signatures. In this study, we analyzed GBM radiomic features on concentric shells, which were then mapped onto 2D planes for radiomics analysis. Radiomic features—including shape features, first-order statistics, and texture descriptors (GLCM, GLRLM, GLDM, GLSZM, NGTDM)—were extracted using PyRadiomics from four different regions in GBM: necrotic core, the T1-weighted contrast-enhancing region, the T2/FLAIR hyperintense lesion, and the 2cm peritumoral expansion region. Feature selection was performed using ANOVA F-statistics. Classification was conducted with multiple machine-learning models, including neural networks, logistic regression, random forest, and TPOT. Model interpretability was evaluated through SHAP analysis, clustering analysis, feature significance profiling, and comparison between radiomic patterns and underlying biological processes.

**Results:** Spherical radiomics consistently outperformed conventional 2D and 3D Cartesian radiomics across all prediction tasks. Radiomic-based neural networks achieved performance comparable to logistic regression and surpassed other tested models. The best framework reached an AUC of 0.85 for MGMT, 0.80 for EGFR, 0.80 for PTEN, and 0.83 for survival prediction. GLCM-derived features were identified as the most informative predictors. Radial transition analysis using the Mann-Whitney U-test demonstrate that transition slopes between T1-weighted contrast-enhancing and T2/FLAIR hyperintense lesion regions, as well as between T2/FLAIR hyperintense lesion and a 2cm peritumoral expansion region, are significantly associated with biomarker status. Furthermore, the observed radiomic changes along the radial direction closely reflected known biological characteristics.

**Conclusion:** Radiomic features extracted on the spherical surfaces at varying radial distances to the GBM tumor centroid are better correlated with important tumor molecular markers and patient survival than the conventional Cartesian analysis. The novel approach improves the performance of non-invasive prediction of key tumor biomarkers and patient outcomes. The work indicates the value of performing quantitative imaging analysis on the manifolds consistent with tumor growth and evolution.

## KEYWORDS

Glioblastoma · Radiogenomic · Spherical Radiomics · Machine Learning · Tumor Heterogeneity

# INTRODUCTION

Glioblastoma (GBM) is the most aggressive and lethal primary brain tumor in adults, accounting for approximately 16% of all primary brain and central nervous system neoplasms, with an age-adjusted incidence rate of 3.2 per 100,000 population [1]. Although GBMs occur almost exclusively in the brain, they can also arise in the brainstem, cerebellum, and spinal cord [2]. Among primary gliomas, 61% occur in the cerebral lobes, most commonly in the frontal (25%), temporal (20%), parietal (13%), and occipital (3%) regions [3]. GBM arises from glial cells—the supportive cells of the central nervous system—and is characterized by rapid proliferation, diffuse infiltration, and resistance to standard therapies. Despite advances in surgical resection, radiotherapy, and chemotherapy, median survival remains only 14–18 months [4].

One major contributor to this poor prognosis is the profound molecular and phenotypic heterogeneity of GBM at genetic, cellular, and radiographic levels [5]. Intertumoral heterogeneity refers to differences in molecular, genetic, or imaging features between tumors from different patients, whereas intratumoral heterogeneity describes diversity within a single tumor mass. GBM frequently exhibits marked intratumoral heterogeneity, which influences treatment response, tumor progression, and overall outcomes [6]. This heterogeneity is evident in the spatial distribution of proliferative, hypoxic, and necrotic regions, as well as in the molecular landscape, where biomarker expressions such as MGMT and EGFR can vary across different tumor subregions [7]. Such diversity has direct clinical implications, including in precise diagnosis, identification of actionable driver mutations, and clinical trial design [8]. Intratumoral heterogeneity can be categorized at the molecular [9,10], cellular [11], and tissue [12] levels, but also poses significant challenges for tissue sampling and biopsy-based molecular testing [13,14].

Given the limitations of invasive brain biopsies in capturing tumor heterogeneity, radiogenomics has emerged as a powerful noninvasive method to correlate quantitative imaging features with underlying genomic profiles [15]. By capturing whole tumor phenotypes that reflect biological processes such as cellularity, angiogenesis, and necrosis [16], radiogenomics can provide complementary insights to molecular testing. Furthermore, radiomics features extracted from 2D or 3D multimodal imaging have been used to predict MGMT promoter methylation [17,18], EGFRvIII mutation [19,20], PTEN [21], and survival status [22] with varying accuracies.

Traditional radiomics approaches—whether based on 2D slices or entire 3D tumor volumes—are based on Cartesian coordinates, which are not intrinsic to the tumor growth pattern. Mathur et al. [23] recently showed that GBM intratumoral heterogeneity demonstrates a strong and statistically significant dependence on the distance to the tumor centroid. For example, the classical and mesenchymal cells are more centrally located vs. the more peripherally located proneuronal and neural subtypes. Also, as shown by Greenwald et al. [24], GBM comprises both disorganized and structured regions, with the structured regions exhibiting a five-layered spatial organization primarily driven by hypoxia. These layers include the hypoxia/necrosis core (layer 1, necrotic core region), a hypoxia-associated layer (layer 2, T1-weighted contrast-enhancing region), an angiogenic response layer (layer 3, T1-weighted contrast-enhancing region), a malignant neurodevelopmental layer (layer 4, T2 hyperintense lesion region), and finally the brain parenchyma (layer 5, 2 cm peritumoral expansion region). These geometrical patterns are consistent with tumor radial growth and evolution. However, existing radiomics analysis based on the Cartesian coordinates is insensitive to the radially evolutionary patterns, which requires a new approach based on the spherical coordinates.

In this study, we propose **spherical radiomics** that extracts features from concentric shells of increasing radii from the tumor centroid to enhance the structural representation of image heterogeneity that is consistent with tumor evolution.

# OVERVIEW OF RESEARCH APPROACH

An overview of this research work is summarized and described in **Figure** 1. The tumor masks for medical images are generated using an ensemble model consisting of prior BraTS challenge-winning segmentation algorithms and then manually corrected by trained radiologists and approved by 2 expert reviewers [25]. Then, spherical radiomics are extracted from each tumor region and further used for developing machine learning (ML) models for molecular and survival status prediction.

In this study, we utilized the dataset UCSF-PDGM, which is a publicly available cohort that includes imaging and molecular data from 501 glioblastoma patients in TCIA (299 valid GBM patients with required molecular and survival information) [25]. This dataset provides diverse imaging protocols and patient demographics, making it well-suited for evaluating model performance under real-world conditions. A demographic description of the patient cohorts is provided in Appendix A. For UCSF-PDGM datasets, we focused on T1-weighted contrast-enhanced (T1CE), FLAIR, and Apparent Diffusion Coefficient (ADC) MRI sequences. T1CE and FLAIR are routinely acquired in standard neuro-oncology workflows and provide complementary information on tumor enhancement and peritumoral edema. ADC, derived from diffusion-weighted imaging, reflects water molecule diffusivity and serves as a surrogate marker for tumor cellularity, offering valuable physiological information that can enhance radiogenomic characterization.

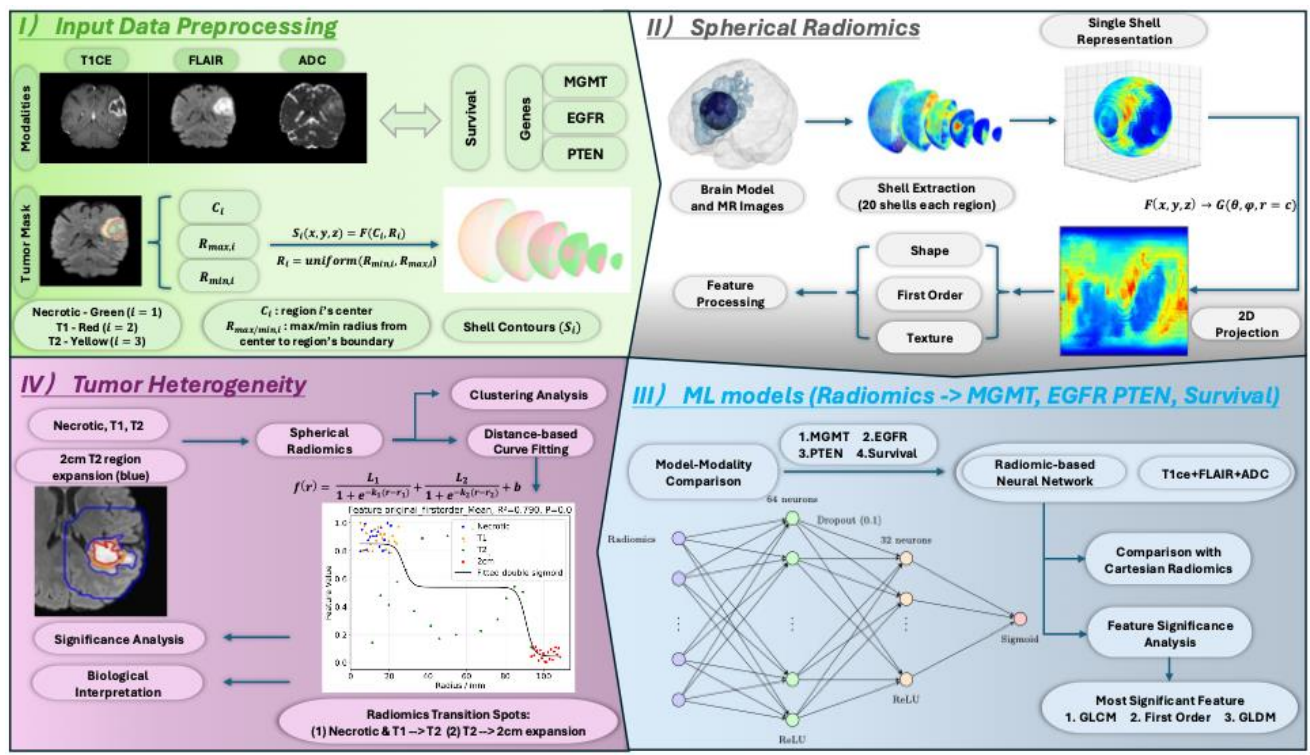

*Figure 1: Study flowchart: (I) Input data consisted of three imaging modalities (T1CE, FLAIR, and ADC) along with molecular markers (MGMT promoter methylation, EGFR and PTEN mutation status) and survival status. Tumor masks were divided into three subregions: necrotic core, enhancing tumor (T1), and edema/infiltrative region (T2). Shell masks were generated using the tumor centroid and the minimum and maximum radii encompassing each subregion. (II) Imaging intensities were interpolated onto the shell masks and projected onto 2D planes. Radiomic features, including shape features from the 3D tumor region, first-order statistics, and multiple texture categories from projected 2D planes, were then extracted from each plane. An ANOVA F-test was applied to select the most discriminative features. (III) Multiple machine learning models and modality combinations were tested. The neural network framework and logistic regression model integrating all modalities achieved the highest predictive performance. Feature importance analysis further revealed that GLCM-based features were the most predictive for different molecular and survival statuses. (IV) To characterize the radiomic gradient beyond the visible tumor boundary, we generated a 2 cm isotropic expansion of the T2 region. Spherical radiomic features extracted from this peritumoral region, together with those from the original tumor regions, were modeled as a function of radius (distance from the tumor center). The transition slope from the tumor core to the surrounding tissue emerged as a significant parameter, reflecting underlying biological processes. Furthermore, clustering analysis highlighted the advantage of spherical radiomics in capturing and explaining tumor heterogeneity.*

# RESULTS

## *Prediction model*

In this study, we implemented a fully connected three-layer neural network to predict MGMT promoter methylation, EGFR mutation status, PTEN mutation status, and patient survival status from extracted radiomic features. We utilized a median survival of 15-month progression-free survival as the efficacy endpoint of therapy trials [26]. The network architecture comprised two hidden layers with 64 and 32 neurons, respectively, each activated with the ReLU function, followed by a sigmoid-activated output layer for binary classification. To reduce the risk of overfitting, a dropout layer with a rate of 0.1 was applied after the first hidden layer, randomly deactivating a fraction of neurons during training and thereby promoting better generalization. Model training employed the binary cross-entropy loss function. Both Adaptive Moment Estimation (Adam) and stochastic gradient descent (SGD) optimizers yielded comparable performance; in this work, we adopted SGD with a learning rate of 0.1 and a decay parameter of 0.001.

To ensure robust evaluation, we performed a 5-fold cross-validation strategy, which allowed model assessment across multiple independent data splits and reduced potential bias. The predictive performance of the neural network was benchmarked against several baseline methods, including logistic regression with ridge regularization, random forest with 100 estimators, and the Tree-based Pipeline Optimization Tool (TPOT) [27], configured with a generation size of 10 and a population size of 20. TPOT, an AutoML framework based on genetic programming, automatically optimizes machine learning pipelines by selecting and combining models and preprocessing steps. Across evaluations, the proposed neural network had comparable performance with logistic regression and demonstrated superior accuracy and stability compared to random forest and TPOT.

## *Influence for number of features*

As outlined in the research overview, we evaluated both mono- and multimodal MR radiomics and subsequently applied predictive modeling to assess the performance of the neural network (NN) relative to logistic regression (LR), random forest (RF), and TPOT. For this study, the top 400 features were selected based on overall predictive performance for gene biomarker and survival status prediction. **Figure** 2 illustrates the impact of feature number on NN-based prediction of MGMT promoter methylation and EGFR mutation.

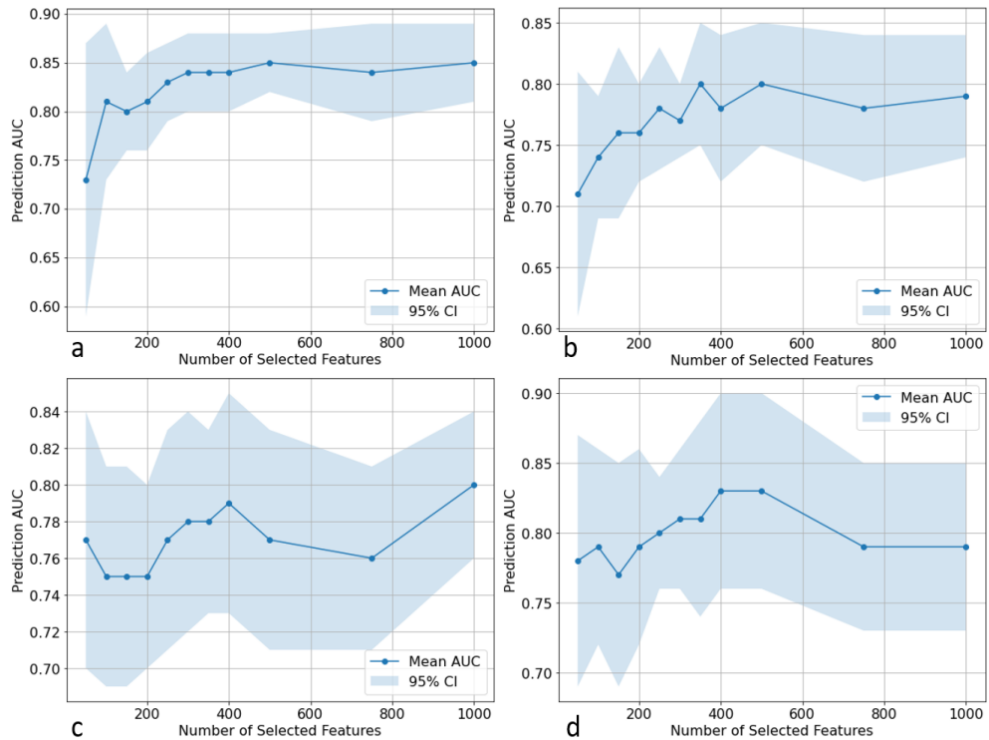

*Figure 2: Number of features' influence on neural network's prediction AUC (band denotes 95% confidence interval): (a) MGMT promoter methylation prediction AUC (b) EGFR mutation prediction AUC (c) PTEN mutation prediction AUC (d) Survival prediction AUC.*

### *Prediction accuracy among different radiomic feature sets*

**Figure** 3 summarizes the predictive performance of four radiomic feature sets—2D radiomics, 3D radiomics, spherical radiomics, and spherical radiomics augmented with tumor mask shape features—across all four modeling algorithms. For all prediction tasks, spherical radiomics consistently outperformed conventional 2D and 3D feature sets. Among the algorithms, neural network (NN) and logistic regression (LR) demonstrated higher accuracy than random forest (RF) and TPOT, underscoring their superior capacity to model high-dimensional radiogenomic data. Notably, LR achieved performance comparable to NN when applied to spherical radiomics; however, its accuracy declined relative to NN for MGMT and EGFR prediction when restricted to conventional 2D or 3D features, whereas it exceeded NN for PTEN and survival prediction under those same feature constraints. **Figure** 4 illustrates the ROC curves for the different models in the first fold of the K-fold cross-validation. The curves visually corroborate the numerical findings, showing that NN and LR consistently achieved more favorable sensitivity–specificity trade-offs than RF and TPOT in spherical radiomics–based predictions. The ROC curves for the remaining folds are provided in Appendix B.

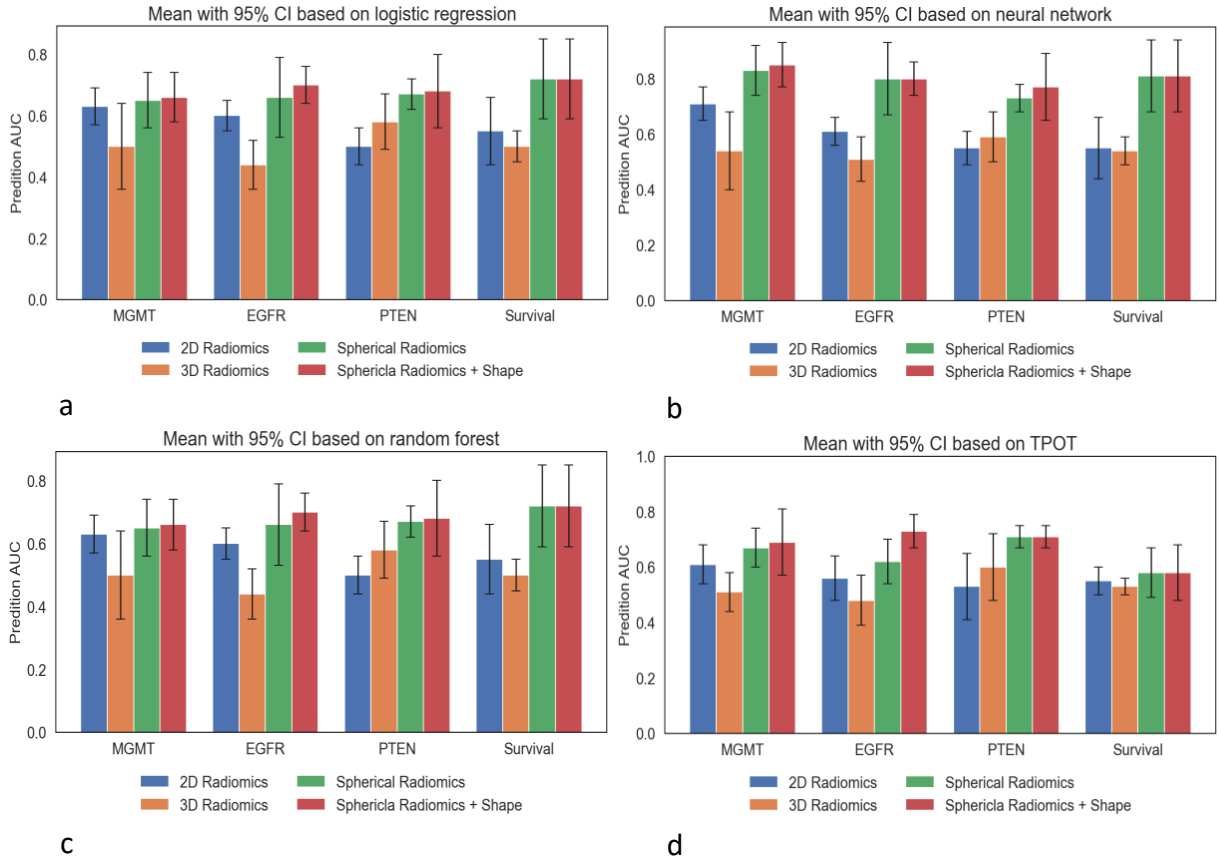

Figure 3: Prediction AUCs for different feature extraction methods and machine learning algorithms using all modalities: (a) Prediction based on logistic regression; (b) Prediction based on neural network; (c) Prediction based on random forest; (d) Prediction based on TPOT.

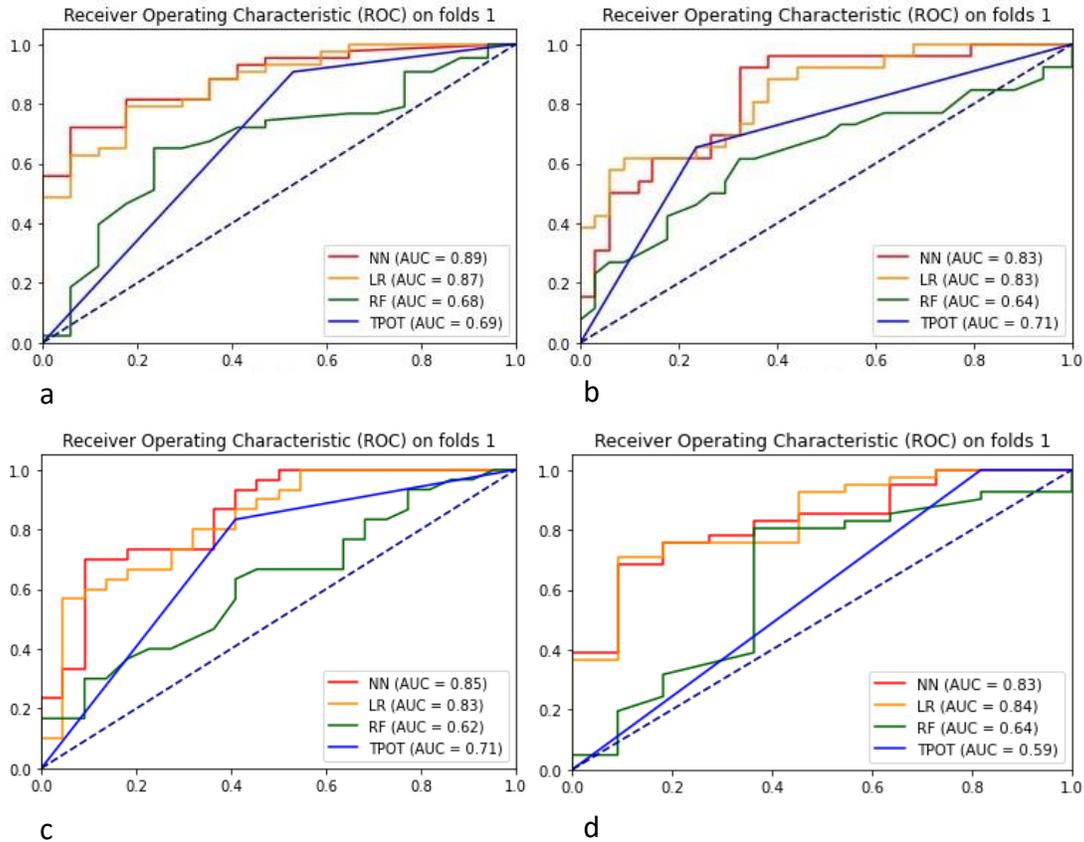

*Figure 4: ROC curves at fold 1 for different machine learning algorithms regarding (X-axis is False Positive Rate and Y-axis True Positive Rate) (a) MGMT prediction; (b) EGFR prediction; (c) PTEN prediction; (d) Survival status prediction.*

### *Clustering comparison between spherical and Cartesian space radiomics*

The benefits of performing radiomic extraction in spherical rather than Cartesian coordinates were further assessed using clustering analysis. To align with our top-performing predictive models (logistic regression and neural networks), we employed principal component analysis (PCA) for dimensionality reduction, taking advantage of its linear projection properties. **Figure** 5 illustrates an example in which GLCM-derived features were clustered in 2D and 3D spaces using linear discriminant analysis (LDA) applied to the PCA projections, comparing features extracted from Cartesian and spherical coordinates. The results demonstrate that GLCM features from spherical radiomics exhibit markedly clearer separation between clusters in both 2D and 3D visualizations compared with their Cartesian counterparts. For instance, in the 2D space, spherical radiomics yielded well-defined clusters, whereas Cartesian radiomics displayed only partial clustering at the extremes, with a large mixed region persisting in the center. Clustering quality was further quantified using the silhouette score [28,29], which ranges from -1 to +1, with higher values indicating better alignment of data points within clusters and stronger separation between clusters. **Figure** 6 summarizes the silhouette scores of four representative radiomic feature classes (First Order, GLCM, GLDM, and GLRLM) using linear projection (PCA). Consistently, spherical radiomics achieved higher silhouette scores than Cartesian radiomics, providing a mechanistic explanation for their superior predictive performance and enhanced ability to capture GBM heterogeneity.

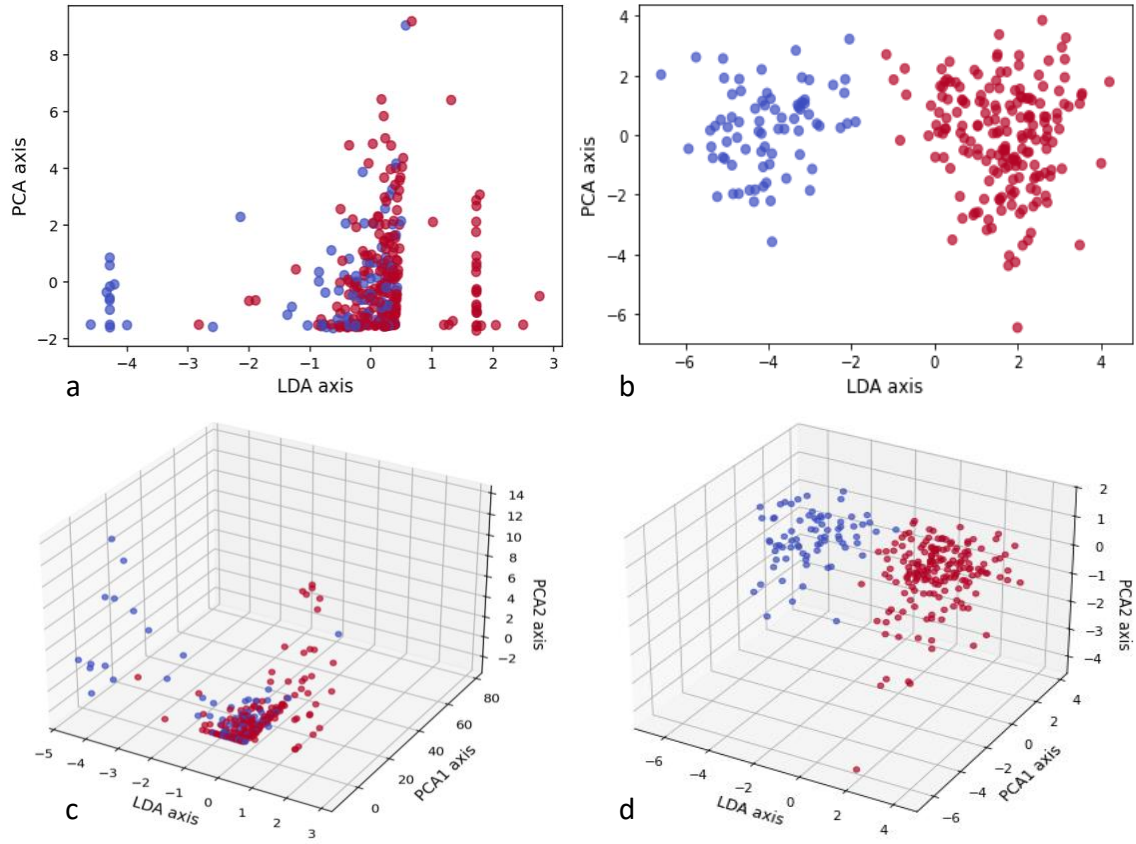

*Figure 5: MGMT label supervised clustering with LDA and PCA projection on GLCM-typed features (blue: patient with negative MGMT label, red: patient with positive MGMT label): (a) 2D space clustering with Cartesian 2D radiomics (silhouette score = 0.07). (b) 2D space clustering with Spherical radiomics (silhouette score = 0.60). (c) 3D space clustering with Cartesian 2D radiomics (silhouette score = 0.06). (d) 3D space clustering with Spherical radiomics (silhouette score = 0.56).*

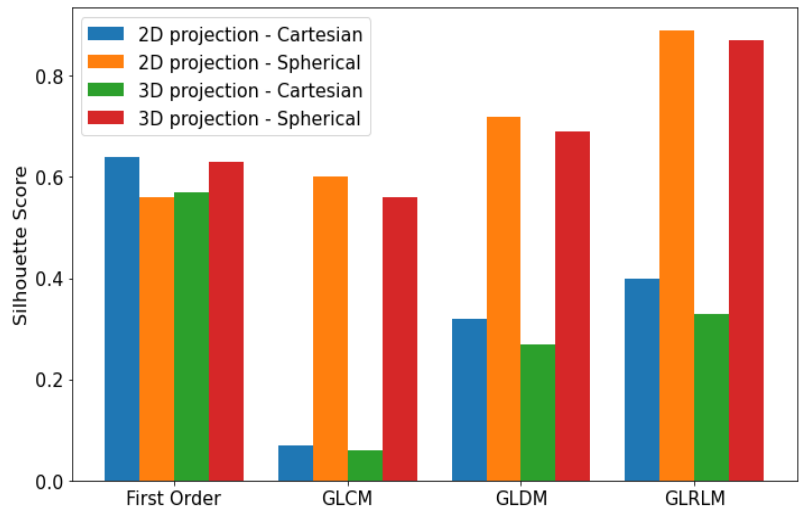

*Figure 6: Silhouette score for top four significant types of radiomics First Order, GLCM, GLDM, GLRLM-typed features) with linear (LDA+PCA) projections onto 2D and 3D spaces.*

*Interpret GBM biological heterogeneity with spherical radiomics*

Single-cell and spatial transcriptomic studies indicate that GBM exhibits five cellular layers radiating from the necrotic core to infiltrated normal brain [24], corresponding to the regions defined in Table 1. We evaluated 93 radiomic features across four MRI-defined zones, using Shapiro–Wilk tests for normality and t-tests or Mann–Whitney U tests for inter-regional differences ($p < 0.05$). As shown in Table 2, the largest differences were observed between the T2 hyperintense region and the 2-cm peritumoral extension zone, while differences between the necrotic core and enhancing T1 region were less pronounced. **Figure** 7 illustrates the first order mean feature, which exhibited significant variation across regions in all three modalities, highlighting the ability of spherical radiomics to capture biologically meaningful heterogeneity in GBM.

*Table 1 Relation between GBM layers and MRI zones*

| GBM layers | Example Dominant Cell States | Functional Interpretation | GBM MRI Zone |
|---|---|---|---|
| Layer 1 | MES hypoxic/necrotic | Hypoxia/Necrotic Core | Necrotic core |
| Layer 2 | MES-like, MES-Ast | Hypoxia-associated States | T1-enhancing |
| Layer 3 | Angiogenetic, Vascular | Angiogenic Response and Immune Hub | T1-enhancing |
| Layer 4 | Astrocyte-like, Neural-progenitor-like | Malignant Neurodevelopment | T2 lesion |
| Layer 5 | Reactive Astrocytes, Neuron | Non-malignant Brain Cells | 2cm expansion, parenchyma |

*Table 2 Percentages of significant difference radiomics between different GBM regions*

| Modalities | Averaged percentage of patients having significant difference for individual radiomics | | | Percentage of radiomics having significant difference with at least one patient | | |
|---|---|---|---|---|---|---|
| | Necrotic vs. T1 | T1 vs. T2 | T2 vs. 2cm | Necrotic vs. T1 | T1 vs. T2 | T2 vs. 2cm |
| *T1CE* | 13.2% | 41.2% | 79.4% | 100% | 100% | 100% |
| *FLAIR* | 18.0% | 50.3% | 77.8% | 100% | 100% | 100% |
| *ADC* | 11.8% | 50.3% | 79.6% | 100% | 100% | 100% |

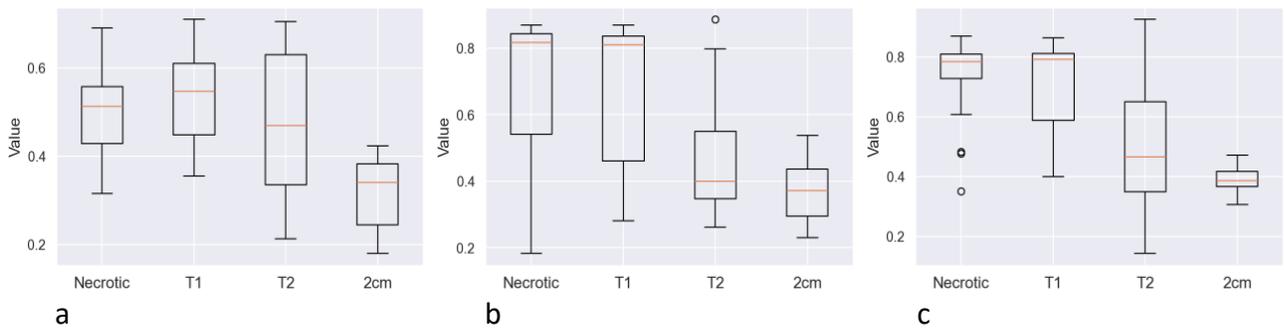

*Figure 7: First order mean boxplot between different pairs of regions described with 20 shells after normalization: (a) T1CE; (b) FLAIR; (c) ADC.*

These results reveal that GBM radiomics exhibit both a radial transition among different tumor sub-volumes and intra-spherical layer heterogeneity. A steeper radial transition in MGMT-unmethylated tumors (negative) was observed compared to MGMT-methylated tumors in T1CE images. Such trends would have been masked in Cartesian radiomics.

### *Tumor heterogeneity described by spherical radiomics*

In addition to analyzing radiomic features on individual shells, we examined tumor radiographic heterogeneity as a function of distance. For each radiomic feature, the observed profile was modeled using a double sigmoid function. This function, defined in Equation 1, integrates two sigmoid components to capture both increasing and decreasing trends, or two-stage transitions, within the feature distribution. $\{L_i\}$ represented the amplitudes of sigmoid function, $\{x_i\}$ indicated the transition point, $\{k_i\}$ described the slope of the transition and $b$ was the constant offset term.

$$f(x) = \frac{L_1}{1+e^{-k_1(x-x_1)}} + \frac{L_2}{1+e^{-k_2(x-x_2)}} + b \tag{1}$$

Fitting the double sigmoid function to various radiomic features revealed that it effectively characterizes two distinct transitions in feature intensity as a function of radial distance from the tumor center. The first transition typically occurred at the boundary between the necrotic and contrast-enhancing tumor core (T1) and the adjacent T2/FLAIR hyperintense region, reflecting a shift from dense tumor cellularity and necrosis to infiltrative edema. The second transition was observed near the periphery of the T2 abnormality, extending into the clinically defined 2-cm expansion zone, and corresponded to an additional change in radiomic intensity and heterogeneity. Although often radiographically subtle, this region is known to harbor infiltrative tumor cells and is routinely encompassed within surgical or radiation treatment margins. **Figure** 8 illustrates an example of a double sigmoid fit for the first order mean feature in patients stratified by MGMT promoter methylation status.

We performed univariate statistical analysis using the Mann–Whitney U test to evaluate the discriminative power of seven radiomic fitting parameters with respect to MGMT promoter methylation, EGFR and PTEN mutations, as well as survival status across different radiomic features. Following FDR correction ($p < 0.05$), approximately 6% of parameters demonstrated significant associations with MGMT promoter methylation, while 15% were significantly associated with EGFR mutation. Among these, the slope parameters k1 and k2, which characterize the steepness of radiomic transitions across regions, accounted for 21% of MGMT-associated features and 47% of EGFR-associated features. Notably, 71% of the significant slope parameters were k1, underscoring its potential sensitivity to genomic status. These results suggest that spatial radiomic transitions—particularly those quantified by k1—may capture underlying molecular heterogeneity and hold promise as imaging biomarkers. **Figure** 8 illustrates the differences in transition slope k1 between MGMT promoter methylation groups for the first order mean feature, while **Figure** 9 visualizes an example of the distinct k1 values observed in two representative patients with differing MGMT promoter methylation status.

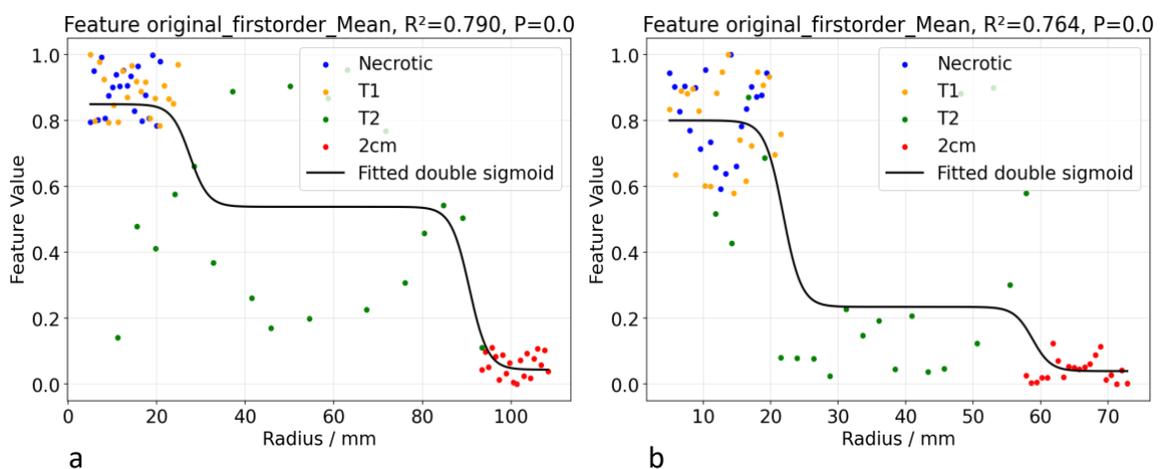

*Figure 8: Double sigmoid fit for first order mean feature in T1CE: (a) A patient with negative MGMT methylation; (b) A patient with positive MGMT methylation. Radiomic feature values in different regions are represented in different colors (blue, yellow, green and red). $R^2$ represents the coefficient of determination and measures the percentage of variance in the data explained by the model. P-value measures the statistical significance of the fitted relationship, where p<0.05 means the fitted relationship is highly statistically significant.*

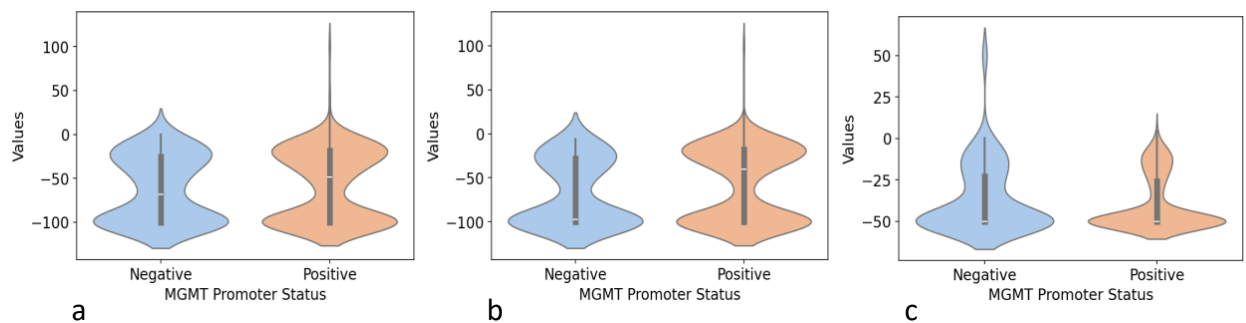

*Figure 9: Violin plot of the transition slope k1 of first order mean feature between negative MGMT promoter methylation (blue) and positive MGMT promoter methylation (orange) for different modalities (Mann-Whitney U test): (a) T1CE (p=0.42); (b) FLAIR (p=0.04); (c) ADC (p=0.87). The central white in black color bar represents the median of the data, and the shape shows the probability density of the data.*

### *Prediction accuracy across different algorithms and modalities based on spherical radiomics*

**Figure** 10 summarizes the prediction AUCs for MGMT, EGFR, PTEN, and survival status across different algorithms and modality combinations using spherical radiomics. Across all modalities and model types, neural network (NN) and logistic regression (LR) consistently outperformed random forest (RF) and TPOT. For example, in predicting MGMT promoter methylation with all three modalities, the NN achieved the highest mean AUC of 0.85 [95% CI: 0.80–0.89], followed by LR (0.84), RF (0.66), and TPOT (0.69). DeLong testing confirmed that NN significantly outperformed RF (p = 0.01, 95% CI: [−0.02, 0.05]) and TPOT (p = 0.05, 95% CI: [−0.08, 0.19]); however, its performance was not significantly different from LR (p = 0.47, 95% CI: [0.18, 0.75]).

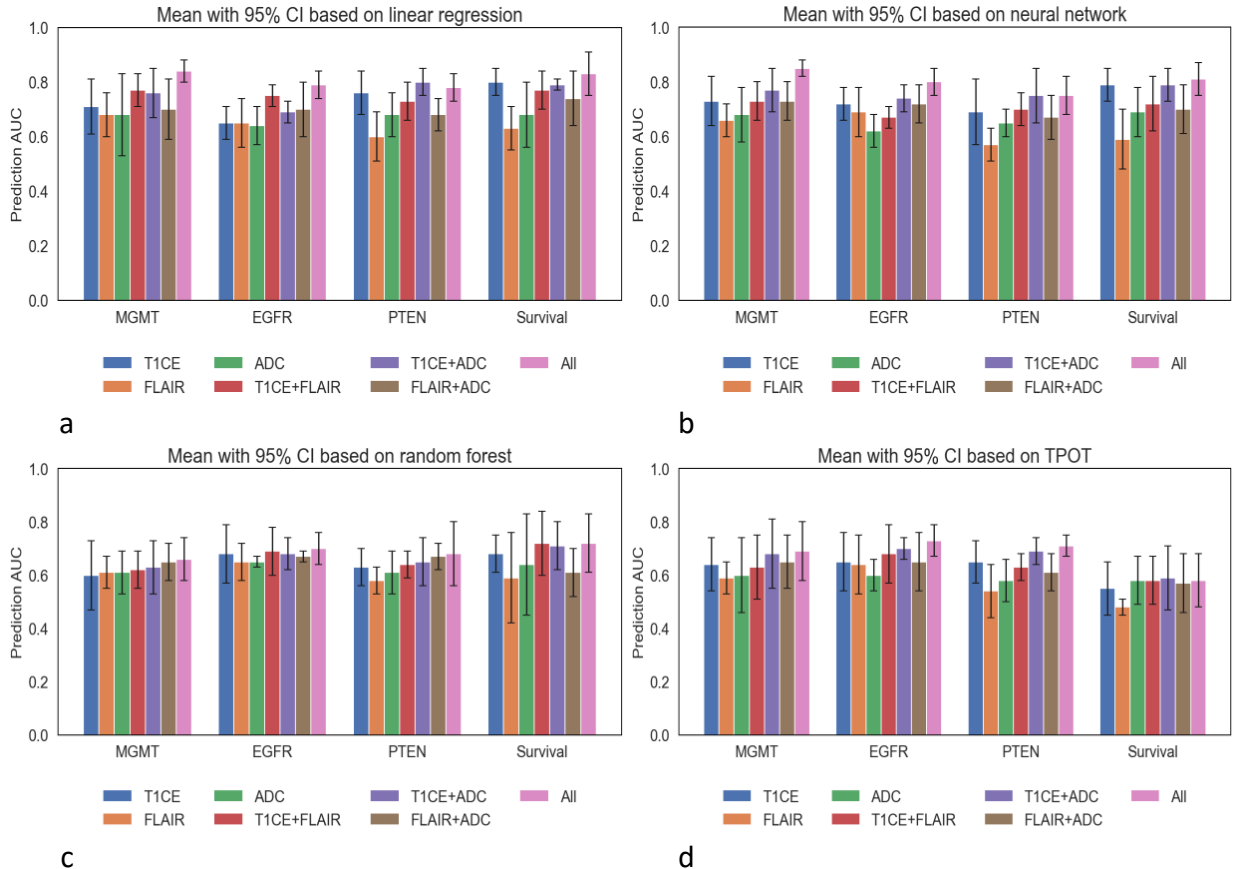

*Figure 10: Prediction AUCs for different modality combinations with different machine learning algorithms: (a) Prediction based on linear regression; (b) Prediction based on neural network; (c) Prediction based on random forest; (d) Prediction based on TPOT.*

### *Feature significance analysis*

To interpret the trained models and identify the most predictive radiomic biomarkers, we employed SHapley Additive exPlanations (SHAP) [30], which quantifies the contribution of each input feature to model predictions at the individual case level and enables ranking of features by overall importance. For illustration, we examined models predicting MGMT promoter methylation and EGFR mutation status. The top 10 most influential features for MGMT and EGFR predictions are shown in **Figure** 11(I) and (II), while the top features for PTEN and survival predictions are presented in **Figure** 12(I) and (II). Among feature categories, gray-level co-occurrence matrix (GLCM) features were particularly prominent, consistently surpassing other categories. For instance, GLCM-based features achieved an AUC of 0.77 for MGMT and 0.68 for EGFR prediction (**Figure** 11(III)). Moreover, their relative representation increased when comparing the original feature pool to the selected significant features: for MGMT and EGFR, GLCM features rose from 25.6% of the original set to 33.4% and 37.2% of the selected features, respectively (**Figure** 11(IV)). Similar patterns were observed for PTEN and survival prediction (**Figure** 12(III) and (IV)). In addition to GLCM features, other categories—including first-order statistics and gray-level dependence matrix (GLDM) features—contributed meaningfully. In summary, GLCM, first-order, and GLDM features achieved competitive AUCs and comprised a notable proportion of the significant predictors, highlighting their complementary predictive value. These results indicate that GLCM features drive model performance, while integrating multiple feature types enhances predictive characterization of molecular and survival outcomes.

## I. Top 10 radiomics for MGMT prediction

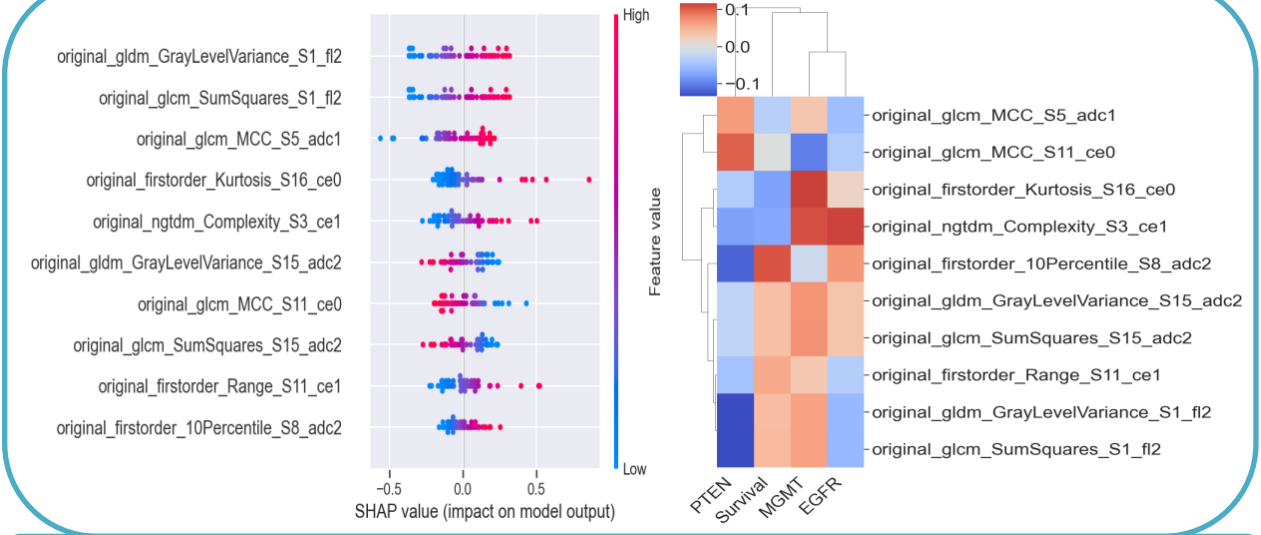

## II. Top 10 radiomics for EGFR prediction

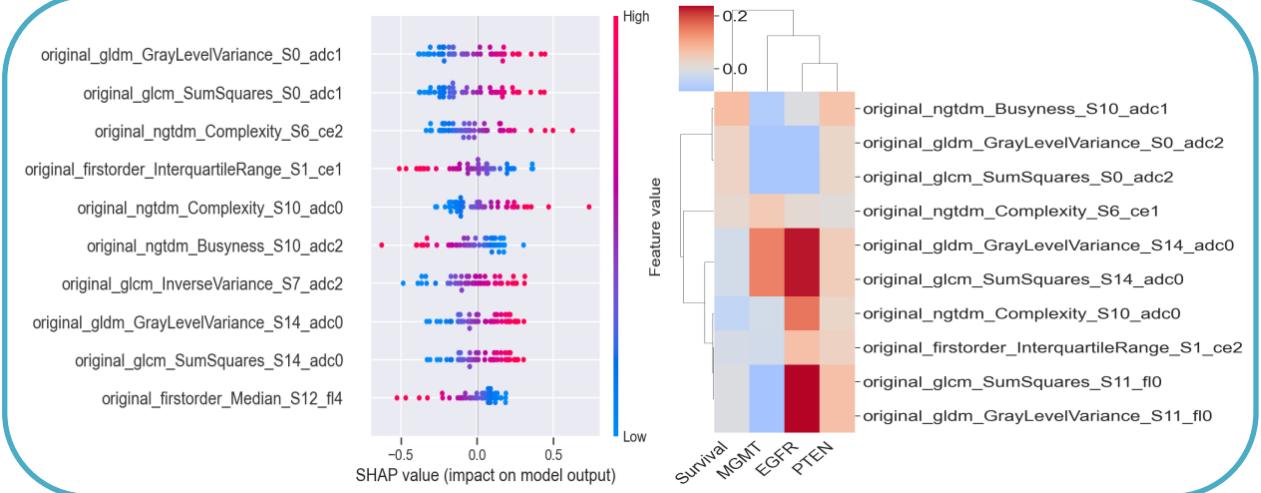

## III. Prediction AUC with different radiomics

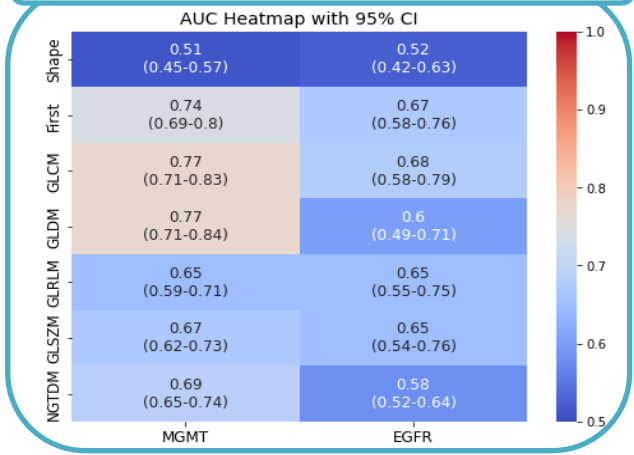

## IV. Percentage of different radiomics

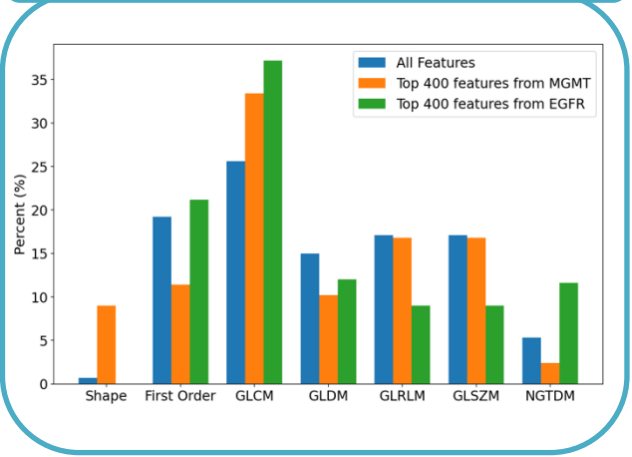

Figure 11: Significant feature analysis: (I) SHAP analysis with hierarchical clustering for MGMT prediction; (II) SHAP analysis with hierarchical clustering for EGFR prediction; (III) Heatmap of prediction AUC across different radiomic feature categories; (IV) Distribution of radiomic feature types among all features (blue), the top 400 features ranked by MGMT prediction (orange), and the top 400 features ranked by EGFR prediction (green).

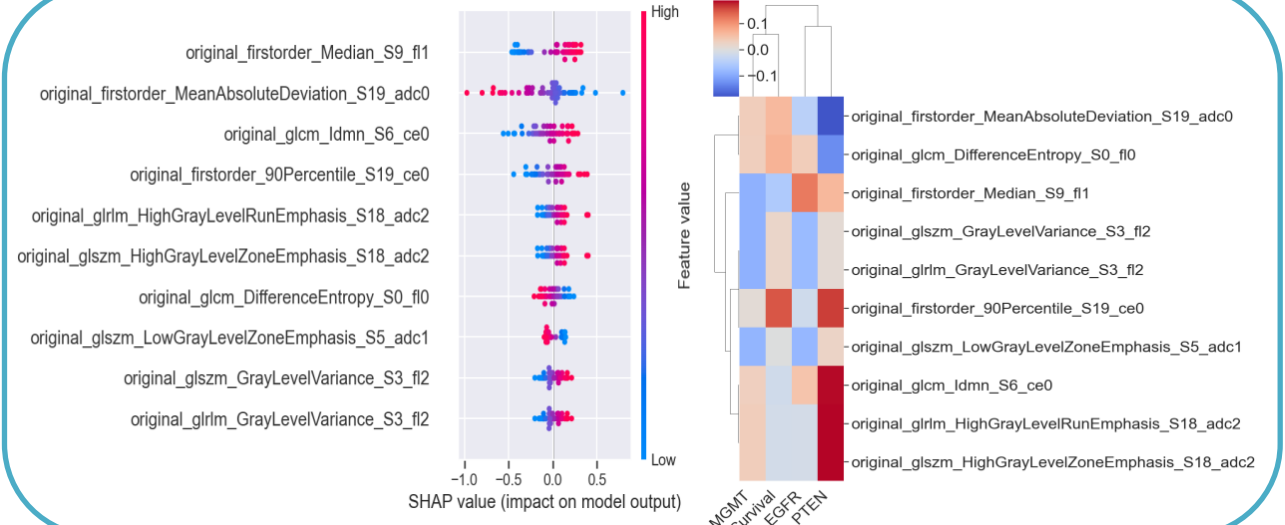

I. Top 10 radiomics for PTEN prediction

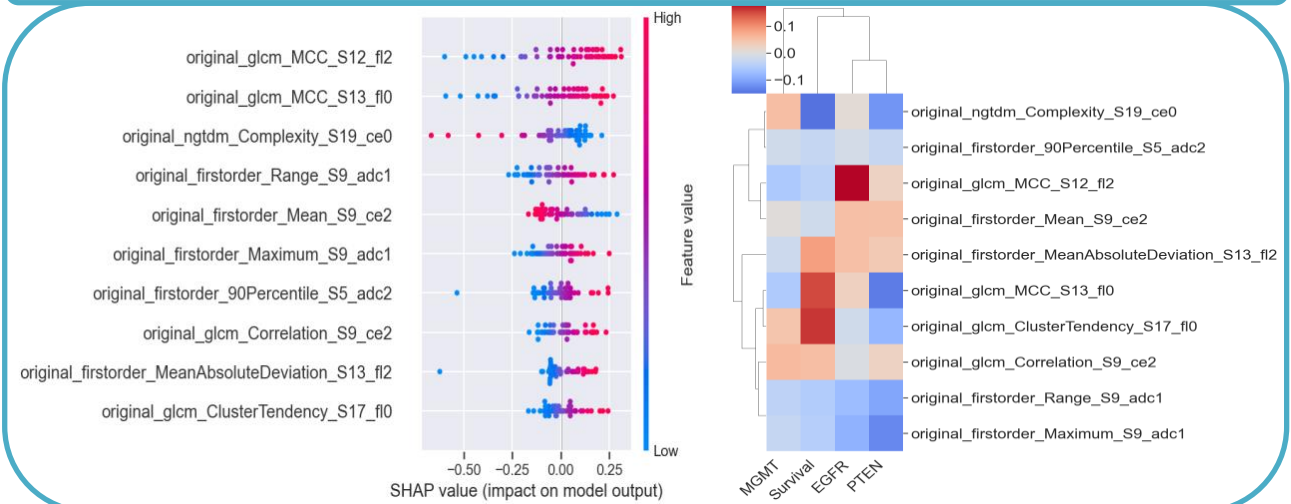

II. Top 10 radiomics for Survival prediction

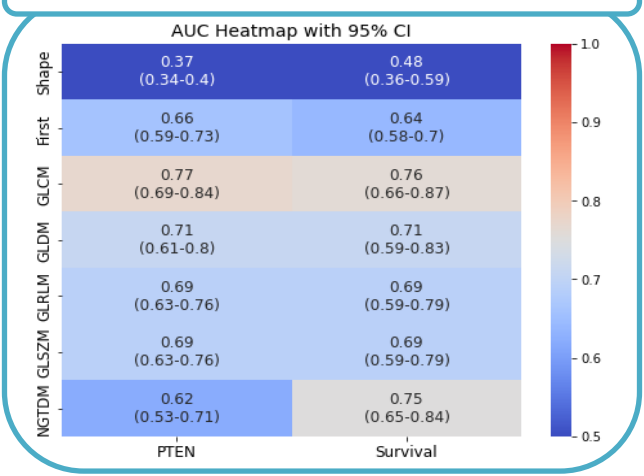

III. Prediction AUC with different radiomics

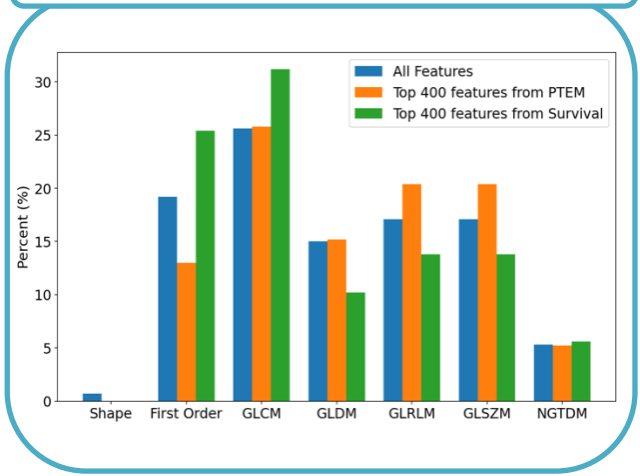

IV. Percentage of different radiomics

*Figure 12: Significant feature analysis: (I) SHAP analysis with hierarchical clustering for PTEN prediction; (II) SHAP analysis with hierarchical clustering for Survival prediction; (III) Heatmap of prediction AUC across different radiomic feature categories; (IV) Distribution of radiomic feature types among all features (blue), the top 400 features ranked by PTEN prediction (orange), and the top 400 features ranked by survival prediction (green).*

## DISCUSSION

The ability to non-invasively predict molecular markers such as MGMT, EGFR, and PTEN, as well as patient survival, has important clinical implications. MGMT is a DNA repair enzyme that removes alkyl groups from the O6 position of guanine [31,32]. While protective in normal cells, MGMT also shields tumor cells from alkylating agents such as temozolomide (TMZ) [31]. Its activity is consumed during repair of TMZ-induced lesions, and intracellular levels correlate with chemoresistance [33]. Promoter methylation silences MGMT expression, reducing repair capacity and increasing tumor sensitivity to TMZ; thus, MGMT promoter methylation is an established predictive and prognostic biomarker in GBM [34,35]. EGFR is another gene frequently altered in GBM, with amplification and the constitutively active mutant EGFRvIII being among the most common events [36,37]. These alterations promote tumor initiation, proliferation, invasion, and therapeutic resistance through oncogenic pathways such as PI3K/AKT [38]. Clinically, EGFR overexpression and EGFRvIII have been linked to aggressive phenotypes, poor prognosis, and treatment resistance in some studies [39,40], although other reports show no clear association with survival [41,42]. PTEN, one of the most frequently altered tumor suppressors in GBM, negatively regulates PI3K signaling. Its loss or mutation results in aberrant pathway activation, driving tumor growth, invasion, and therapy resistance [43]. PTEN function may be disrupted by loss of heterozygosity, mutations, or epigenetic silencing such as DNA methylation [44,45]. Because of its central role in GBM biology, PTEN status is increasingly recognized as a potential prognostic biomarker and a factor guiding patient management and therapeutic strategies [45].

For MGMT, accurate preoperative prediction of promoter methylation status could identify patients most likely to benefit from alkylating chemotherapy (e.g., temozolomide). Similarly, prediction of EGFR alterations, including EGFRvIII, may guide enrollment in targeted therapy trials and provide prognostic information. PTEN, a critical tumor suppressor gene, has been linked to radiation sensitivity and resistance to anti-angiogenic therapies [21]. Some studies have associated PTEN loss with poor survival [46], whereas others reported no significant correlation [47], underscoring the complexity of its role in GBM biology. Beyond molecular markers, accurate imaging-based prediction of survival status could provide clinicians with valuable insight into disease trajectory and support more personalized treatment planning [48]. Early prediction of patient survival is critical to improve the balance between the aggressiveness of intervention and the patients' quality of life.

In addition to their direct predictive value, imaging-based approaches complement biopsy-based testing, which is limited by sampling error, sparse time points, and intratumoral heterogeneity. Our spherical radiomics framework offers several further clinical advantages. First, by capturing distinct radiomic signatures across tumor compartments, this approach can support more personalized surgical planning by improving margin delineation that reflects true tumor infiltration. Second, its capacity to characterize peritumoral heterogeneity may help refine radiotherapy target volumes, enhancing local control while reducing exposure to healthy tissue. Finally, the interpretability and spatial structure of spherical radiomic features offer opportunities for integration into clinical decision-support tools, enabling more precise radiogenomic stratification and personalized treatment strategies.

Nevertheless, predicting genomic status directly from medical images remains challenging, with prior studies reporting modest and variable levels of performance. For example, Sasaki et al. achieved a prediction accuracy of 0.67 for MGMT prediction [49], Haubold et al. achieved an AUC of 0.74 [50], and Xi et al. reported an accuracy of 0.8 [51], while the MICCAI 2021 challenge reported a best AUC of 0.62 for MGMT. These results are consistent with our own observations using standard Cartesian-space radiomics. For EGFR prediction, Rathore et al. reported a best AUC of 0.80 using the Hospital of the University of Pennsylvania dataset with Cartesian-space radiomics [52]. Similarly, Kazerooni et al. reported AUCs of 0.68 for EGFR and 0.73 for PTEN using conventional MRI [53], while Hu et al. achieved accuracies of 0.75 and 0.69 for EGFR and PTEN, respectively, with 3D Cartesian-space radiomics [13].

Beyond genomic prediction, radiomics has also been applied to survival modeling: Bae et al. reported an AUC of 0.65 for survival status prediction using Cartesian-space radiomics [54], and Shboul et al. achieved an accuracy of 0.68 when employing a radiomics feature-guided neural network [55]. It is worth noting that the aforementioned performances were achieved under varying experimental conditions, making direct comparison difficult. For instance, customized cutoff thresholds were used to boost performance, but such approaches substantially limit the generalizability of machine learning models across datasets from different institutions [56].

In this study, we demonstrated that spherical radiomics achieves statistically significant and substantially higher prediction accuracy than Cartesian-space radiomics for genomic features (EGFR, MGMT, PTEN) and survival *under identical training and testing conditions*. To understand what contributes to the improved prediction performance, it is essential to discuss the foundation of radiomics. The values of radiomics features are driven by the voxel-level heterogeneity in the image intensity [57]. Radiogenomics assumes that there is an underlying correlation between genomics and the formation of tumors and surrounding normal tissues, which is reflected in medical images. However, despite the large panel of radiomics features that have been devised, they do not by default interpret the high-level and global architecture that is the hallmark of GBM [24]. The current study manually encoded a simple tumor global architecture consistent with the native tumor development into the radiomics study. In other words, radiomics' ability to perform quantitative imaging texture analysis needs to be augmented by structural decomposition of the tumor architecture. The performance of radiomics thus depends on the methods of decomposition. The observation is further substantiated by the clustering analysis that spherical radiomics features are more distinctly separated by the biomarker and survival statuses. The transition of the radiomics values also follows the intrinsic radial structure of the GBM, which would have been missed by the standard radiomics analysis performed on a Cartesian grid. Interestingly, the gradient of transition seems to be an important parameter closely correlated with GBM biomarkers such as MGMT methylation status.

The comparative evaluation of predictive models demonstrated an overall advantage of the neural network (NN) relative to the baseline approaches. The use of neural network–based radiomics models has been increasingly reported in oncology, offering the ability to capture nonlinear and high-dimensional interactions that may be overlooked by conventional methods. For example, such models have been successfully applied to prediction tasks in glioblastoma recurrence [58], colorectal cancer [59,60], and tongue cancer [61]. Compared with image-based deep learning frameworks such as convolutional neural networks (CNNs) [62,63], the NN-based radiomics approach offers notable advantages. By leveraging handcrafted features extracted via radiomics, it bypasses the need for large-scale training datasets while still achieving competitive or superior predictive accuracy. This efficiency makes it especially suitable for studies with limited cohort sizes, such as those commonly encountered in glioblastoma research.

Nonetheless, LR remains widely adopted in radiogenomic studies due to its interpretability and robustness [64,65,66]. In our study, although NN achieved marginally higher predictive accuracy, the current sample size and variability limit definitive conclusions about its advantage over LR, warranting validation in larger cohorts. Beyond the comparison of prediction models, this study also emphasizes the added value of integrating spherical radiomic features derived from multiple imaging modalities to enhance predictive accuracy. By leveraging complementary information across modalities, spherical multimodal analysis captures distinct but interrelated aspects of tumor biology, including ADC, T1CE, and FLAIR. This integrative strategy provides a more comprehensive characterization of tumor phenotype than any single modality. The benefits of multimodality radiomics have also been demonstrated in previous studies, where conventional 2D or 3D radiomic features were applied for predicting genetic alterations [13], patient survival [67], and treatment response [68,69]. Our results extend these findings by showing that the combination of multimodality input with spherical spatially resolved features yields further improvements, suggesting that both the choice of feature representation and the integration of complementary imaging data are critical for advancing radiogenomic prediction.

Moreover, our feature significance analysis revealed that gray-level co-occurrence matrix (GLCM) features were particularly prominent, consistently outperforming other feature categories across SHAP importance ranking, hierarchical clustering, and predictive accuracy analyses. This finding indicates that texture heterogeneity captured by GLCM features reflects biologically meaningful variations in tumor microstructure that may correspond to underlying genomic alterations. GLCM features quantify second-order voxel intensity statistics, representing spatial dependencies and intra-tumoral heterogeneity—attributes frequently linked to tumor aggressiveness and genetic instability. As shown in Figures 11 and 12, GLCM Sum of Squares (Variance) and Maximal Correlation Coefficient (MCC) repeatedly emerged among the top contributors. MCC characterizes the complexity and dependency of gray-level relationships, while Sum of Squares measures the dispersion of co-occurrence probabilities around the mean [29]. Their consistent prominence suggests that genomic alterations and survival status may be associated with heightened texture heterogeneity and structural complexity within the tumor.

Our findings align with previous studies showing that GLCM-derived metrics possess superior discriminative power. For instance, GLCM-based features extracted in Cartesian space have been reported to achieve higher prediction accuracy than other feature types in applications such as brain tumor classification [70], cervical cancer [71], and breast cancer [72]. Our analysis extends their conclusions by demonstrating that the predictive strength of GLCM features remains robust within spherical geometry.

The study has several limitations that we wish to discuss here.

1. Generalizability: The study was performed on data from a single institution. Due to the lack of available external data, including the same imaging sequences, biomarkers, and clinical information, testing of its robustness and generalizability will be performed in future studies.
2. Feature stability and biological interpretation: Spherical radiomics offers a novel approach to feature extraction by capturing peritumoral patterns that may reflect underlying biological processes. However, the use of a uniform spherical shell is still a simplification of the complex tumor microenvironment, particularly in glioblastoma, where diffusion, perfusion, and infiltration patterns are highly heterogeneous and anisotropic. Although spherical radiomics shows a significantly better correlation with several key GBM biomarkers and patient survival, the mechanistic understanding of the correlation remains challenging.
3. A major driver of the current study is spatially encoded transcriptomics. However, such information is unavailable for the current study patient cohort. Therefore, the spherical radiomics features were only correlated with biomarkers that are not specific to a geometric location of the tumor. A more in-depth analysis based on spatially encoded transcriptomics samples registered to multi-parametric MR images will likely yield additional insight into the spherical radiogenomics.

## CONCLUSIONS

In this study, we developed a novel spherical radiomic framework for the prediction of MGMT promoter methylation, EGFR mutation, PTEN mutation, and survival statuses in glioblastoma, integrating features derived from multiple imaging modalities. Compared with conventional 2D and 3D radiomics, spherical radiomics achieves consistently higher predictive accuracy across all modeling approaches. The inclusion of shape features further enhanced performance, highlighting the importance of spatial context in radiogenomic analysis.

## METHODS

### Different tumor regions of GBM

Glioblastoma (GBM) typically exhibits four anatomically and biologically distinct regions identifiable on MRI: the necrotic core, the T1-weighted contrast-enhancing region, the T2/FLAIR hyperintense lesion, and the 2cm expansion region [73]. The necrotic core, usually visualized as a non-enhancing central area, reflects hypoxic tissue death and is associated with aggressive tumor biology and poor prognosis. Surrounding this core is the contrast-enhancing rim on post-contrast T1-weighted images, which indicates regions of high vascular permeability and blood–brain barrier breakdown. This area contains the bulk of proliferative tumor cells and is typically the primary target for surgical resection and radiotherapy. Beyond the enhancing rim lies the T2/FLAIR hyperintense region, which encompasses both infiltrative tumor cells and vasogenic edema. Although this region appears less aggressive radiologically, it is known to harbor microscopic tumor invasion and is frequently implicated in recurrence following treatment.

In addition to these radiographically visible regions, clinical guidelines recommend a 2cm isotropic expansion beyond the visible tumor boundary to define the peritumoral region at risk for microscopic infiltration [74]. This margin accounts for the highly infiltrative nature of GBM, as tumor cells can extend well beyond the enhancing core. The margin is included in the radiotherapy target volume as part of the standard GBM management protocol [75]. In this study, the 2cm expansion mask was generated by applying binary morphological dilation to the lesion mask, using a 3D spherical structuring element scaled according to the voxel dimensions.

Each of these tumor regions provides complementary biological information. By analyzing them separately through radiomic features, we could achieve a more nuanced understanding of intratumoral heterogeneity and explore its association with key molecular alterations such as $O^6$-methylguanine-DNA methyltransferase (MGMT) promoter methylation, Epidermal growth factor receptor (EGFR) mutation, phosphatase and tensin homolog (PTEN) mutation, as well as survival status.

### *Spherical radiomics and feature extraction*

In this study, we introduced a novel radiomic framework termed *spherical radiomics*, specifically designed to capture the radial growth pattern of glioblastoma (GBM). We segmented the tumor and surrounding volumes into a series of concentric 3D shells—thin, non-overlapping layers that evolve outward from the geometric center of the tumor toward its margin. Radiomic features were then extracted independently from each shell, enabling a localized, layer-wise characterization of tumor heterogeneity. This spherical decomposition allowed us to analyze spatial gradients in texture, intensity, and shape features that may correspond to various molecular statuses and survival statuses.

1. Shell contour generation

First, shells were generated as spherical shapes centered at the center of the tumor region. Specifically, for each tumor subregion, we generated N uniformly spaced shells. That is, assuming $r_{max}$ represented the maximum radius of the sphere shell that covers the tumor region's outer boundary and $r_{min}$ represented the minimum radius of the spherical shell that covers the tumor region's inner boundary, the radius of the *i–th* spherical shell was determined as $r_i = r_{min} + i \frac{r_{max} - r_{min}}{N}$. In this study, we chose *N* = 20, and 8000 points were sampled from each shell. An example of the extracted shell from the tumor center, as well as contour sampling and projection on shell surfaces, is shown in Figure 13.

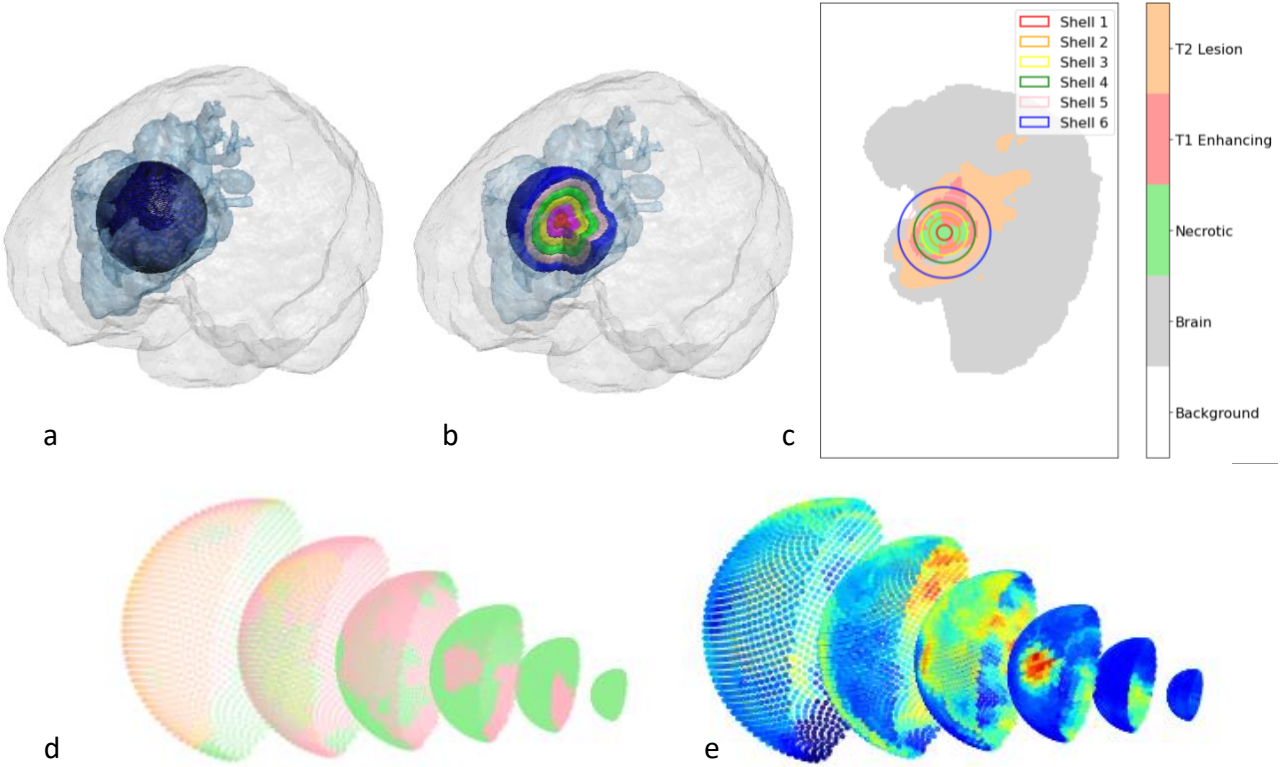

*Figure 13: Example of spherical shell extraction in the tumor region with six shells: (a) 3D full view showing the spatial relationship between brain (light gray), tumor (light blue), and generated shells. (b) Cutout view of the spherical shells. (c) 2D slice through the tumor center. (d, e) Half spherical shell visualization with 2000 sampling points: (d) shell mask originated from tumor center, where green represents necrotic region, light red represents T1 region and orange represents T2 region (e) T1CE intensities on the corresponding shell mask.*

2. Shell contour mapping onto 2D spherical coordinates

To extract the corresponding radiomic features using the standard radiomics formula, we mapped the shell contour onto a 2D Cartesian plane. A spherical shape is described using spherical coordinates r (radius), φ (azimuthal angle or longitude, in the range [−π, π]), and $\theta$ (spherical angle or colatitude, in the range [0, π]).

On each shell surface, we queried the corresponding pixel value of each sampled point with Equation 2. **Figure** 14 shows the examples of several shell contour mappings of T1CE, FLAIR, and ADC images in the T1-enhancing region. Compared to traditional 2D planes across the tumor that are insensitive to the radial transition, the projected spherical surfaces capture the radial evolution of imaging characteristics from the necrotic core to the peritumoral region. Example of mapping results including all 20 shell layers at individual region (necrotic, T1 enhancing and T2 lesion region) is presented in Appendix C.

$$\begin{bmatrix} x \\ y \\ z \end{bmatrix} = r \cdot \begin{bmatrix} sin\varphi \cdot cos\theta \\ sin\varphi \cdot sin\theta \\ cos\varphi \end{bmatrix} \quad (2)$$

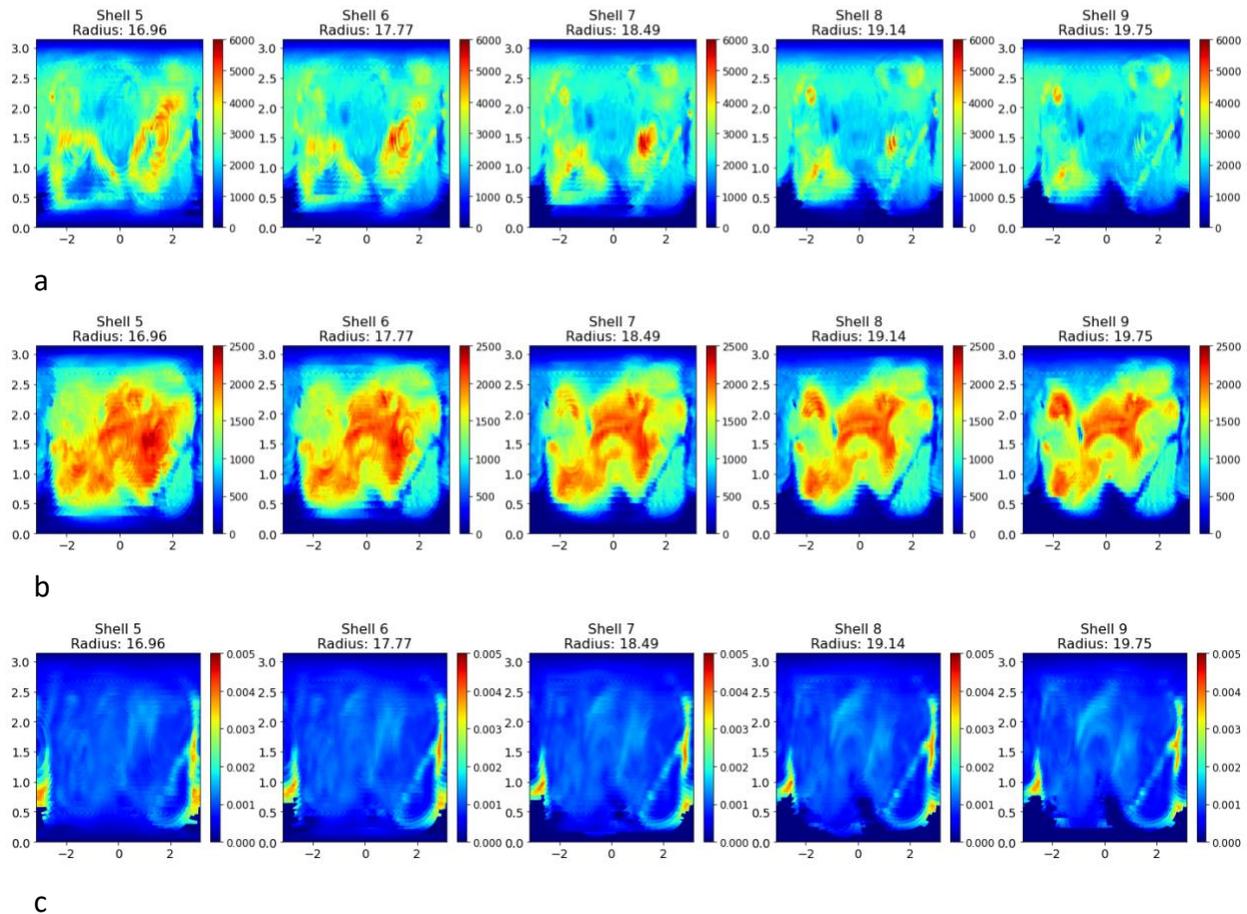

*Figure 14: Example of shell contour mappings in T1 enhancing region (shell 5 to shell 9) from: (a) T1CE image; (b) FLAIR image; (c) ADC image.*

3. Feature collection and selection

In this study, we extracted radiomic features from individual MRI modalities (T1CE, FLAIR, and ADC) as well as from their combinations to explore the benefits of multimodal integration using PyRadiomics [76]. To identify the most informative features for predicting genetic mutations, we applied the SelectKBest algorithm, a univariate feature selection method that ranks features based on statistical relevance to the target variable. Specifically, we employed the Analysis of Variance (ANOVA) F-test as the scoring function, which assesses the degree of variance between and within groups defined by the target classes b. Features with higher F-scores indicate stronger discriminatory power and are more likely to be predictive. This approach provided an efficient means of reducing feature dimensionality and mitigating overfitting by filtering out irrelevant or noisy features.

**RESOURCE AVAILABILITY**

The entire framework can be found on our GitHub page: https://github.com/Isaac0047/Shell_Radiomics. git. The raw data required to reproduce the findings presented in the paper are available to download from https://www.cancerimagingarchive.net/collection/ucsf-pdgm/.

Requests for further information and resources should be directed to and will be fulfilled by the corresponding author, Ke Sheng (Ke.Sheng@ucsf.edu).

## AUTHOR CONTRIBUTIONS

Conceptualization, H.F. and K.S.; methodology, H.F.; investigation, H.F. and K.S.; writing—original draft, H.F. and K.S.; funding acquisition, K.S.; supervision, K.S.

## DECLARATION OF INTERESTS

The authors declare that they have no known competing financial interests or personal relationships that could have appeared to influence the work reported in this paper.

## DECLARATION OF GENERATIVE AI AND AI-ASSISTED TECHNOLOGIES

During the preparation of this work, the authors used ChatGPT in order to polish the readability of this manuscript. After using this tool or service, the authors reviewed and edited the content as needed and take full responsibility for the content of the publication.

## REFERENCES

[1] Ostrom QT, Bauchet L, Davis FG, Deltour I, Fisher JL, Langer CE, Pekmezci M, Schwartzbaum JA, Turner MC, Walsh KM, et al. The epidemiology of glioma in adults: a "state of the science" review. Neuro-oncology. 2014 Jul 1;16(7):896-913.

[2] Davis ME. Glioblastoma: overview of disease and treatment. Clinical journal of oncology nursing. 2016 Oct 1;20(5):S2.

[3] Lovely MP, Stewart-Amidei C, Arzbaecher J, Bell S, Maher ME, Maida M, Mogensen K, Nicolaseau G. Care of the adult patient with a brain tumor. Journal of Neuroscience Nursing. 2014 Dec 1;46(6):367-9.

[4] Weng J, Salazar N. DNA methylation analysis identifies patterns in progressive glioma grades to predict patient survival. International journal of molecular sciences. 2021 Jan 20;22(3):1020.

[5] Friedmann-Morvinski D. Glioblastoma heterogeneity and cancer cell plasticity. Critical Reviews™ in Oncogenesis. 2014;19(5).

[6] Becker AP, Sells BE, Haque SJ, Chakravarti A. Tumor heterogeneity in glioblastomas: from light microscopy to molecular pathology. Cancers. 2021 Feb 12;13(4):761.

[7] Qazi MA, Vora P, Venugopal C, Sidhu SS, Moffat J, Swanton C, Singh SK. Intratumoral heterogeneity: pathways to treatment resistance and relapse in human glioblastoma. Annals of Oncology. 2017 Jul 1;28(7):1448-56.

[8] Young JS, Prados MD, Butowski N. Using genomics to guide treatment for glioblastoma. Pharmacogenomics. 2018 Oct 1;19(15):1217-29.

[9] DeCordova S, Shastri A, Tsolaki AG, Yasmin H, Klein L, Singh SK, Kishore U. Molecular heterogeneity and immunosuppressive microenvironment in glioblastoma. Frontiers in immunology. 2020 Jul 17;11:1402.

[10] Eder K, Kalman B. Molecular heterogeneity of glioblastoma and its clinical relevance. Pathology & Oncology Research. 2014 Oct;20(4):777-87.

[11] Eisenbarth D, Wang YA. Glioblastoma heterogeneity at single cell resolution. Oncogene. 2023 Jun 30;42(27):2155-65.

[12] Gallaher JA, Massey SC, Hawkins-Daarud A, Noticewala SS, Rockne RC, Johnston SK, Gonzalez-Cuyar L, Juliano J, Gil O, Swanson KR, et al. From cells to tissue: How cell scale heterogeneity impacts glioblastoma growth and treatment response. PLoS computational biology. 2020 Feb 26;16(2):e1007672.

[13] Hu LS, Ning S, Eschbacher JM, Baxter LC, Gaw N, Ranjbar S, Plasencia J, Dueck AC, Peng S, Smith KA, et al. Radiogenomics to characterize regional genetic heterogeneity in glioblastoma. Neuro-oncology. 2017 Jan 1;19(1):128-37.

[14] Hu LS, D'Angelo F, Weiskittel TM, Caruso FP, Fortin Ensign SP, Blomquist MR, Flick MJ, Wang L, Sereduk CP, Meng-Lin K, et al. Integrated molecular and multiparametric MRI mapping of high-grade glioma identifies regional biologic signatures. Nature communications. 2023 Sep 28;14(1):6066.

[15] Mazurowski MA. Radiogenomics: what it is and why it is important. Journal of the American College of Radiology. 2015 Aug 1;12(8):862-6.

[16] Kuo MD, Jamshidi N. Behind the numbers: decoding molecular phenotypes with radiogenomics—guiding principles and technical considerations. Radiology. 2014 Feb;270(2):320-5.

[17] Li ZC, Bai H, Sun Q, Li Q, Liu L, Zou Y, Chen Y, Liang C, Zheng H. Multiregional radiomics features from multiparametric MRI for prediction of MGMT methylation status in glioblastoma multiforme: a multicentre study. European radiology. 2018 Sep;28(9):3640-50.

[18] Kickingereder P, Neuberger U, Bonekamp D, Piechotta PL, Götz M, Wick A, Sill M, Kratz A, Shinohara RT, Jones DT, et al. Radiomic subtyping improves disease stratification beyond key molecular, clinical, and standard imaging characteristics in patients with glioblastoma. Neuro-oncology. 2018 May 18;20(6):848-57.

[19] Akbari H, Bakas S, Pisapia JM, Nasrallah MP, Rozycki M, Martinez-Lage M, Morrissette JJ, Dahmane N, O'Rourke DM, Davatzikos C. In vivo evaluation of EGFRvIII mutation in primary glioblastoma patients via complex multiparametric MRI signature. Neuro-oncology. 2018 Jul 5;20(8):1068-79.

[20] Bakas S, Akbari H, Pisapia J, Martinez-Lage M, Rozycki M, Rathore S, Dahmane N, O'Rourke DM, Davatzikos C. In vivo detection of EGFRvIII in glioblastoma via perfusion magnetic resonance imaging signature consistent with deep peritumoral infiltration: The φ-index. Clinical Cancer Research. 2017 Aug 15;23(16):4724-34.

[21] Chen H, Lin F, Zhang J, Lv X, Zhou J, Li ZC, Chen Y. Deep learning radiomics to predict PTEN mutation status from magnetic resonance imaging in patients with glioma. Frontiers in Oncology. 2021 Oct 4;11:734433.

[22] Babaei Rikan S, Sorayaie Azar A, Naemi A, Bagherzadeh Mohasefi J, Pirnejad H, Wiil UK. Survival prediction of glioblastoma patients using modern deep learning and machine learning techniques. Scientific Reports. 2024 Jan 29;14(1):2371.

[23] Mathur R, Wang Q, Schupp PG, Nikolic A, Hilz S, Hong C, Grishanina NR, Kwok D, Stevers NO, Jin Q, et al. Glioblastoma evolution and heterogeneity from a 3D whole-tumor perspective. Cell. 2024 Jan 18;187(2):446-63.

[24] Greenwald AC, Darnell NG, Hoefflin R, Simkin D, Mount CW, Castro LN, Harnik Y, Dumont S, Hirsch D, Nomura M, et al. Integrative spatial analysis reveals a multi-layered organization of glioblastoma. Cell. 2024 May 9;187(10):2485-501.

[25] Calabrese E, Villanueva-Meyer JE, Rudie JD, Rauschecker AM, Baid U, Bakas S, Cha S, Mongan JT, Hess CP. The University of California San Francisco preoperative diffuse glioma MRI dataset. Radiology: Artificial Intelligence. 2022 Oct 5;4(6):e220058.

[26] Stupp R, Hegi ME, Mason WP, Van Den Bent MJ, Taphoorn MJ, Janzer RC, Ludwin SK, Allgeier A, Fisher B, Belanger K, et al. Effects of radiotherapy with concomitant and adjuvant temozolomide versus radiotherapy alone on survival in glioblastoma in a randomised phase III study: 5-year analysis of the EORTC-NCIC trial. The lancet oncology. 2009 May 1;10(5):459-66.

[27] Olson RS, Bartley N, Urbanowicz RJ, Moore JH. Evaluation of a tree-based pipeline optimization tool for automating data science. InProceedings of the genetic and evolutionary computation conference 2016 2016 Jul 20 (pp. 485-492).

[28] Shutaywi M, Kachouie NN. Silhouette analysis for performance evaluation in machine learning with applications to clustering. Entropy. 2021 Jun 16;23(6):759.

[29] Rousseeuw PJ. "Silhouettes: a graphical aid to the interpretation and validation of cluster analysis." *Journal of computational and applied mathematics* 20 (1987): 53-65.

[30] Lundberg SM, Lee SI. A unified approach to interpreting model predictions. Advances in neural information processing systems. 2017;30.

[31] Pegg AE. Repair of O6-alkylguanine by alkyltransferases. Mutation Research/Reviews in Mutation Research. 2000 Apr 1;462(2-3):83-100.

[32] Thon N, Kreth S, Kreth FW. Personalized treatment strategies in glioblastoma: MGMT promoter methylation status. OncoTargets and therapy. 2013 Sep 27:1363-72.

[33] Gerson SL. MGMT: its role in cancer aetiology and cancer therapeutics. Nature Reviews Cancer. 2004 Apr 1;4(4):296-307.

[34] Weller M, Stupp R, Reifenberger G, Brandes AA, Van Den Bent MJ, Wick W, Hegi ME. MGMT promoter methylation in malignant gliomas: ready for personalized medicine?. Nature Reviews Neurology. 2010 Jan;6(1):39-51.

[35] Mansouri A, Hachem LD, Mansouri S, Nassiri F, Laperriere NJ, Xia D, Lindeman NI, Wen PY, Chakravarti A, Mehta MP, et al. MGMT promoter methylation status testing to guide therapy for glioblastoma: refining the approach based on emerging evidence and current challenges. Neuro-oncology. 2019 Feb 14;21(2):167-78.

[36] Xu H, Zong H, Ma C, Ming X, Shang M, Li K, He X, Du H, Cao L. Epidermal growth factor receptor in glioblastoma. Oncology letters. 2017 Jul;14(1):512-6.

[37] An Z, Aksoy O, Zheng T, Fan QW, Weiss WA. Epidermal growth factor receptor and EGFRvIII in glioblastoma: signaling pathways and targeted therapies. Oncogene. 2018 Mar 22;37(12):1561-75.

[38] Gan HK, Kaye AH, Luwor RB. The EGFRvIII variant in glioblastoma multiforme. Journal of Clinical Neuroscience. 2009 Jun 1;16(6):748-54.

[39] Li J, Liang R, Song C, Xiang Y, Liu Y. Prognostic significance of epidermal growth factor receptor expression in glioma patients. OncoTargets and therapy. 2018 Feb 7:731-42.

[40] Haas-Kogan DA, Prados MD, Tihan T, Eberhard DA, Jelluma N, Arvold ND, Baumber R, Lamborn KR, Kapadia A, Malec M, et al. Epidermal growth factor receptor, protein kinase B/Akt, and glioma response to erlotinib. Journal of the National Cancer Institute. 2005 Jun 15;97(12):880-7.

[41] Heimberger AB, Suki D, Yang D, Shi W, Aldape K. The natural history of EGFR and EGFRvIII in glioblastoma patients. Journal of translational medicine. 2005 Oct 19;3(1):38.

[42] Hoogstrate Y, Ghisai SA, De Wit M, De Heer I, Draaisma K, Van Riet J, Van De Werken HJ, Bours V, Buter J, Vanden Bempt I, et al. The EGFRvIII transcriptome in glioblastoma: A meta-omics analysis. Neuro-oncology. 2022 Mar 1;24(3):429-41.

[43] Koul D. PTEN signaling pathways in glioblastoma. Cancer biology & therapy. 2008 Sep 1;7(9):1321-5.

[44] Wiencke JK, Zheng S, Jelluma N, Tihan T, Vandenberg S, Tamgüney T, Baumber R, Parsons R, Lamborn KR, Berger MS, et al. Methylation of the PTEN promoter defines low-grade gliomas and secondary glioblastoma. Neuro-oncology. 2007 Jul 1;9(3):271-9.

[45] Han F, Hu R, Yang H, Liu J, Sui J, Xiang X, Wang F, Chu L, Song S. PTEN gene mutations correlate to poor prognosis in glioma patients: a meta-analysis. OncoTargets and therapy. 2016 Jun 13:3485-92.

[46] Garcia-Carracedo D, Turk AT, Fine SA, Akhavan N, Tweel BC, Parsons R, Chabot JA, Allendorf JD, Genkinger JM, Remotti HE, et al. Loss of PTEN expression is associated with poor prognosis in patients with intraductal papillary mucinous neoplasms of the pancreas. Clinical Cancer Research. 2013 Dec 15;19(24):6830-41.

[47] Abdullah JM, Farizan A, Asmarina K, Zainuddin N, Ghazali MM, Jaafar H, Isa MN, Naing NN. Association of loss of heterozygosity and PTEN gene abnormalities with paraclinical, clinical modalities and survival time of glioma patients in Malaysia. Asian Journal of Surgery. 2006 Oct 1;29(4):274-82.

[48] Babaei Rikan S, Sorayaie Azar A, Naemi A, Bagherzadeh Mohasefi J, Pirnejad H, Wiil UK. Survival prediction of glioblastoma patients using modern deep learning and machine learning techniques. Scientific Reports. 2024 Jan 29;14(1):2371.

[49] Sasaki T, Kinoshita M, Fujita K. Radiomics and MGMT promoter methylation for prognostication of newly diagnosed glioblastoma. Sci Rep 2019;9:14435. 10.1038/s41598-019-50849-y.

[50] Haubold J, Hosch R, Parmar V, Glas M, Guberina N, Catalano OA, Pierscianek D, Wrede K, Deuschl C, Forsting M, et al. Fully automated MR based virtual biopsy of cerebral gliomas. Cancers. 2021 Dec 8;13(24):6186.

[51] Xi YB, Guo F, Xu ZL, Li C, Wei W, Tian P, Liu TT, Liu L, Chen G, Ye J, et al. Radiomics signature: A potential biomarker for the prediction of MGMT promoter methylation in glioblastoma. Journal of Magnetic Resonance Imaging. 2018 May;47(5):1380-7.

[52] Rathore S, Mohan S, Bakas S, Sako C, Badve C, Pati S, Singh A, Bounias D, Ngo P, Akbari H, et al. Multi-institutional noninvasive in vivo characterization of IDH, 1p/19q, and EGFRvIII in glioma using neuro-Cancer Imaging Phenomics Toolkit (neuro-CaPTk). Neuro-oncology advances. 2020 Dec 1;2(Supplement_4):iv22-34.

[53] Fathi Kazerooni A, Akbari H, Hu X, Bommineni V, Grigoriadis D, Toorens E, Sako C, Mamourian E, Ballinger D, Sussman R, et al. The radiogenomic and spatiogenomic landscapes of glioblastoma and their relationship to oncogenic drivers. Communications Medicine. 2025 Mar 1;5(1):55.

[54] Bae S, Choi YS, Ahn SS, Chang JH, Kang SG, Kim EH, Kim SH, Lee SK. Radiomic MRI phenotyping of glioblastoma: improving survival prediction. Radiology. 2018 Dec;289(3):797-806.

[55] Shboul ZA, Alam M, Vidyaratne L, Pei L, Elbakary MI, Iftekharuddin KM. Feature-guided deep radiomics for glioblastoma patient survival prediction. Frontiers in neuroscience. 2019 Sep 20;13:966.

[56] Li Y, Liu X, Xu K, Qian Z, Wang K, Fan X, Li S, Wang Y, Jiang T. MRI features can predict EGFR expression in lower grade gliomas: a voxel-based radiomic analysis. European radiology. 2018 Jan;28(1):356-62.

[57] Kim D, Jensen LJ, Elgeti T, Steffen IG, Hamm B, Nagel SN. Radiomics for everyone: a new tool simplifies creating parametric maps for the visualization and quantification of radiomics features. Tomography. 2021 Sep 17;7(3):477-87.

[58] Shim KY, Chung SW, Jeong JH, Hwang I, Park CK, Kim TM, Park SH, Won JK, Lee JH, Lee ST, et al. Radiomics-based neural network predicts recurrence patterns in glioblastoma using dynamic susceptibility contrast-enhanced MRI. Scientific reports. 2021 May 11;11(1):9974.

[59] Chen LD, Li W, Xian MF, Zheng X, Lin Y, Liu BX, Lin MX, Li X, Zheng YL, Xie XY, et al. Preoperative prediction of tumour deposits in rectal cancer by an artificial neural network–based US radiomics model. European radiology. 2020 Apr;30(4):1969-79.

[60] Shi R, Chen W, Yang B, Qu J, Cheng Y, Zhu Z, Gao Y, Wang Q, Liu Y, Li Z, et al. Prediction of KRAS, NRAS and BRAF status in colorectal cancer patients with liver metastasis using a deep artificial neural network based on radiomics and semantic features. American journal of cancer research. 2020 Dec 1;10(12):4513.

[61] Zhong YW, Jiang Y, Dong S, Wu WJ, Wang LX, Zhang J, Huang MW. Tumor radiomics signature for artificial neural network-assisted detection of neck metastasis in patient with tongue cancer. Journal of Neuroradiology. 2022 Mar 1;49(2):213-8.

[62] Calabrese E, Rudie JD, Rauschecker AM, Villanueva-Meyer JE, Clarke JL, Solomon DA, Cha S. Combining radiomics and deep convolutional neural network features from preoperative MRI for predicting clinically relevant genetic biomarkers in glioblastoma. Neuro-Oncology Advances. 2022 Jan 1;4(1):vdac060.

[63] Fukuma R, Yanagisawa T, Kinoshita M, Shinozaki T, Arita H, Kawaguchi A, Takahashi M, Narita Y, Terakawa Y, Tsuyuguchi N, et al. Prediction of IDH and TERT promoter mutations in low-grade glioma from magnetic resonance images using a convolutional neural network. Scientific reports. 2019 Dec 30;9(1):20311.

[64] Hong EK, Choi SH, Shin DJ, Jo SW, Yoo RE, Kang KM, Yun TJ, Kim JH, Sohn CH, Park SH, et al. Radiogenomics correlation between MR imaging features and major genetic profiles in glioblastoma. European radiology. 2018 Oct;28(10):4350-61.

[65] Wijethilake N, Islam M, Ren H. Radiogenomics model for overall survival prediction of glioblastoma. Medical & Biological Engineering & Computing. 2020 Aug;58(8):1767-77.

[66] Kickingereder P, Bonekamp D, Nowosielski M, Kratz A, Sill M, Burth S, Wick A, Eidel O, Schlemmer HP, Radbruch A, et al. Radiogenomics of glioblastoma: machine learning–based classification of molecular characteristics by using multiparametric and multiregional MR imaging features. Radiology. 2016 Dec;281(3):907-18.

[67] Chang K, Zhang B, Guo X, Zong M, Rahman R, Sanchez D, Winder N, Reardon DA, Zhao B, Wen PY, et al. Multimodal imaging patterns predict survival in recurrent glioblastoma patients treated with bevacizumab. Neuro-oncology. 2016 Dec 1;18(12):1680-7.

[68] Verduin M, Compter I, Steijvers D, Postma AA, Eekers DB, Anten MM, Ackermans L, Ter Laan M, Leijenaar RT, van de Weijer T, et al. Noninvasive glioblastoma testing: multimodal approach to monitoring and predicting treatment response. Disease markers. 2018;2018(1):2908609.

[69] Peeken JC, Goldberg T, Pyka T, Bernhofer M, Wiestler B, Kessel KA, Tafti PD, Nüsslin F, Braun AE, Zimmer C, et al. Combining multimodal imaging and treatment features improves machine learning-based prognostic assessment in patients with glioblastoma multiforme. Cancer medicine. 2019 Jan;8(1):128-36.

[70] Dheepak G, Vaishali D. Brain tumor classification: a novel approach integrating GLCM, LBP and composite features. Frontiers in oncology. 2024 Jan 30;13:1248452.

[71] Feng H, Yoshida E, Sheng K. Multi-Modality and Temporal Analysis of Cervical Cancer Treatment Response. arXiv preprint arXiv:2408.13408. 2024 Aug 23.

[72] Elfira E, Priharti W, Rahmawati D. Comparison of GLCM and First Order Feature Extraction Methods for Classification of Mammogram Images. Jurnal Teknokes. 2022 Dec 8;15(4):197-205.

[73] Nelson SJ, Cha S. Imaging glioblastoma multiforme. The Cancer Journal. 2003 Mar 1;9(2):134-45.

[74] Hochberg FH, Pruitt A. Assumptions in the radiotherapy of glioblastoma. Neurology. 1980 Sep;30(9):907-.

[75] Stupp R, Mason WP, Van Den Bent MJ, Weller M, Fisher B, Taphoorn MJ, Belanger K, Brandes AA, Marosi C, Bogdahn U, Curschmann J. Radiotherapy plus concomitant and adjuvant temozolomide for glioblastoma. New England journal of medicine. 2005 Mar 10;352(10):987-96.

[76] Van Griethuysen JJ, Fedorov A, Parmar C, Hosny A, Aucoin N, Narayan V, Beets-Tan RG, Fillion-Robin JC, Pieper S, Aerts HJ. Computational radiomics system to decode the radiographic phenotype. Cancer research. 2017 Nov 1;77(21):e104-7.

# SUPPLEMENTAL INFORMATION

## A PATIENT DEMOGRAPHIC COMPARISON

*Table A Patients demographic comparison of UCSF-PDGM dataset*

| No. of valid GBM patients | No. of men | No. of women | Mean age (y) |
|---|---|---|---|
| 299 | 188 (63%) | 111 (37%) | 62 ± 13 |
| MGMT methylated | EGFR mutated | PTEN mutated | Survived (15-months, 257 patients) |
| 214 (72%) | 129 (43%) | 169 (57%) | 54 (21%) |

## B ROC CURVE FOR DIFFERENT FOLDS

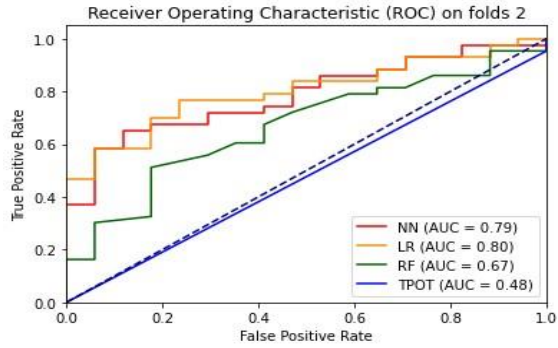
a
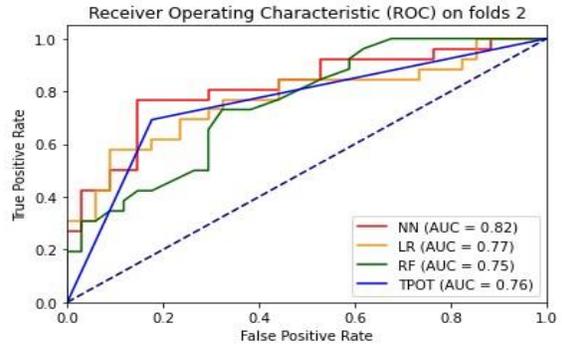
b
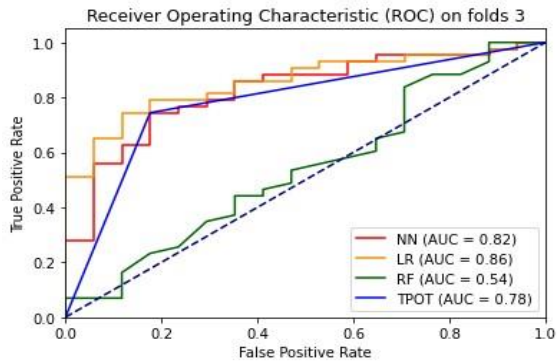
c
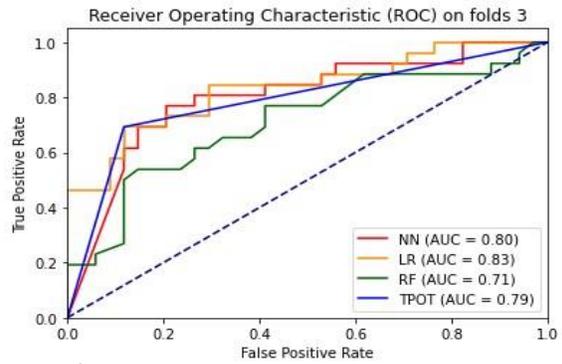
d
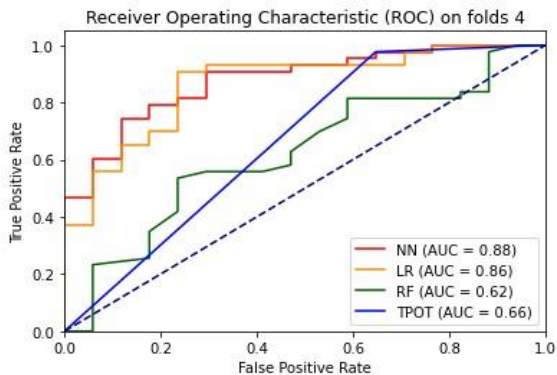
e
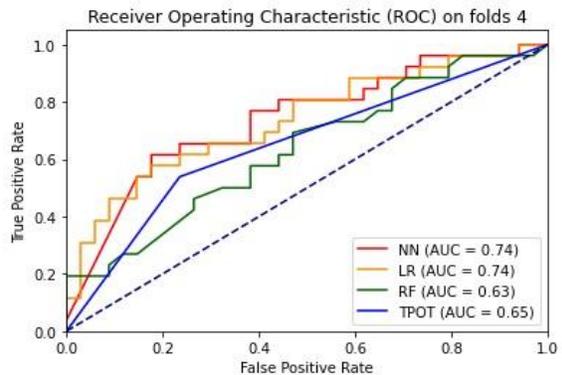
f
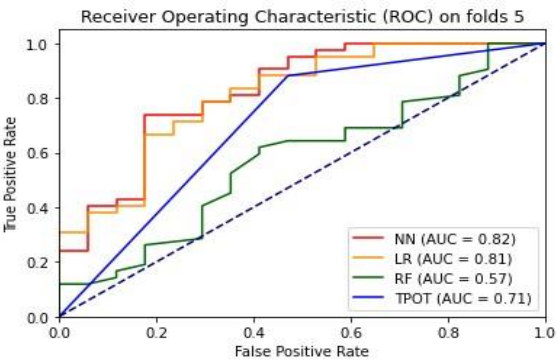
g
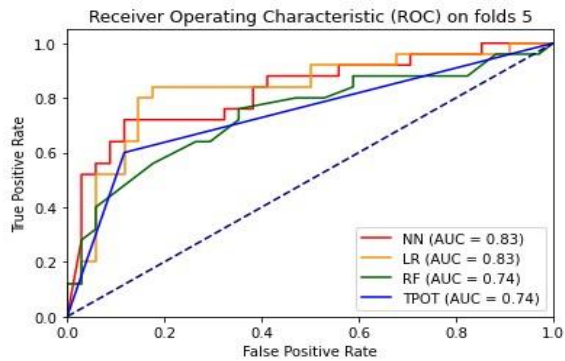
h

*Figure B.1: ROC curve in one fold for (a)(c)(e)(g) MGMT prediction ROC curve on different folds (b) (d) (f) (h) EGFR prediction ROC curve on different folds*

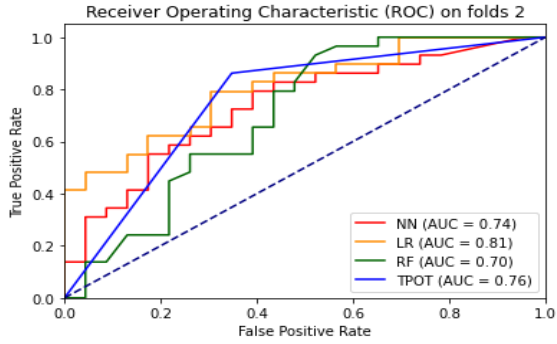
a

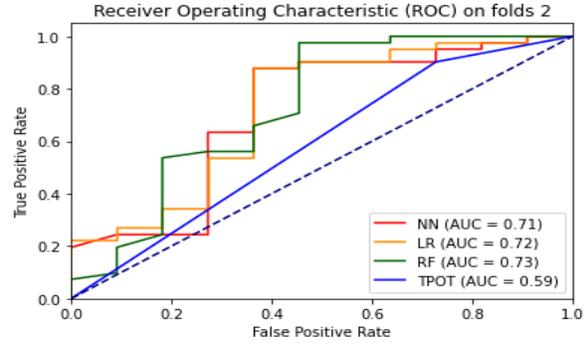
b

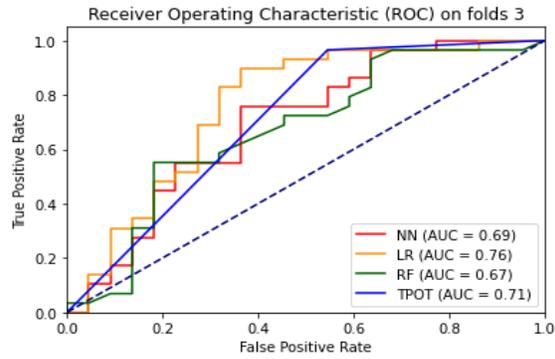
c

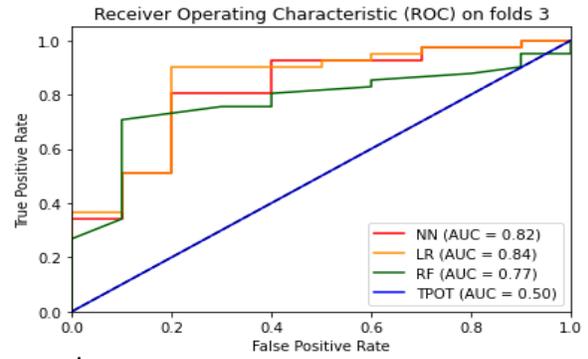
d

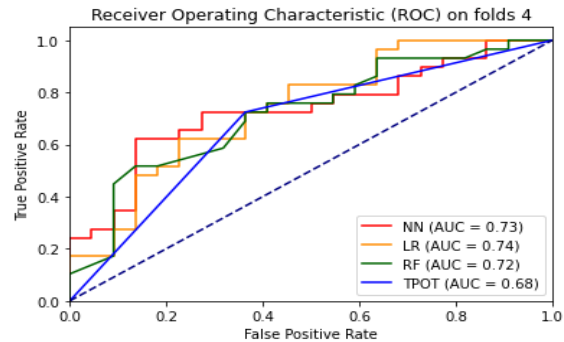
e

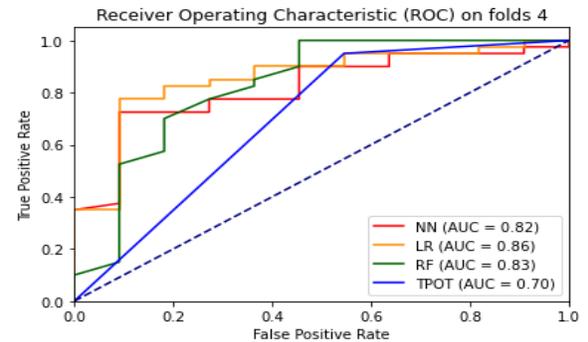
f

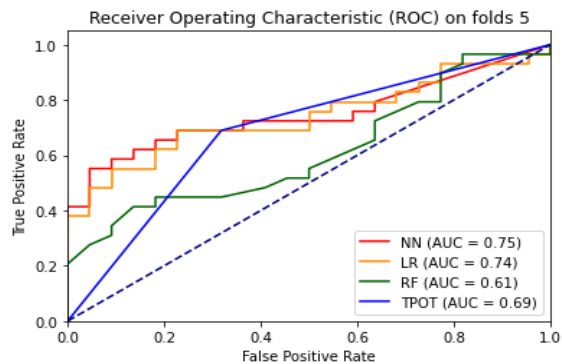
g

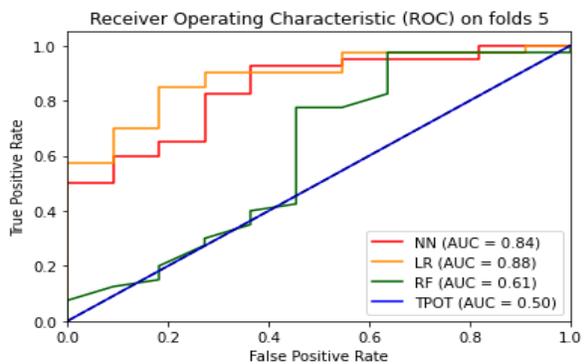
h

*Figure B.2: ROC curve in one fold for (a)(c)(e)(g) PTEN prediction ROC curve on different folds (b) (d) (f) (h) survival prediction ROC curve on different folds*

## C SHELL CONTOUR MAPPINGS

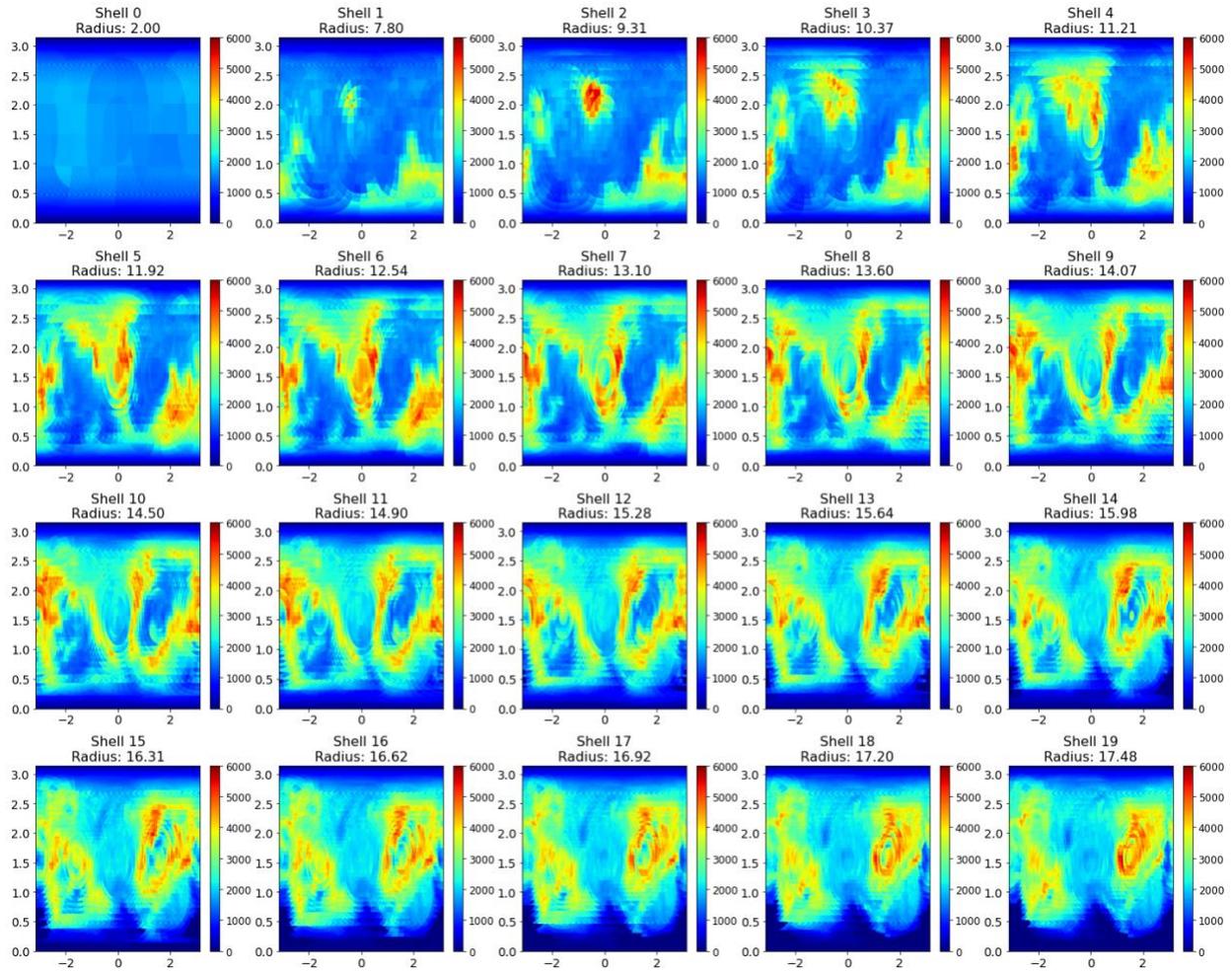

Figure C.1: Shell contour mapping for T1CE in necrotic region

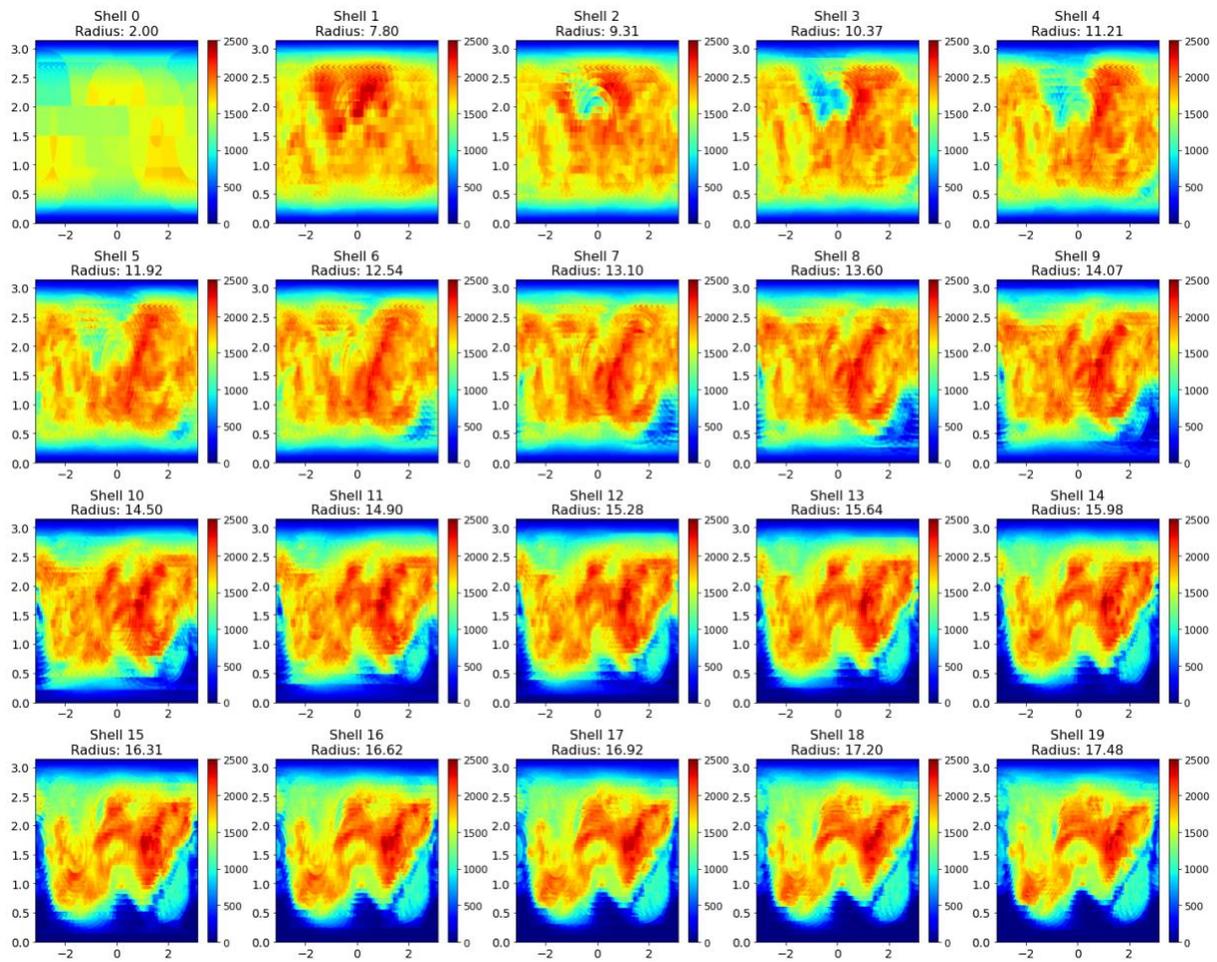

Figure C.2: Shell contour mapping for FLAIR in necrotic region

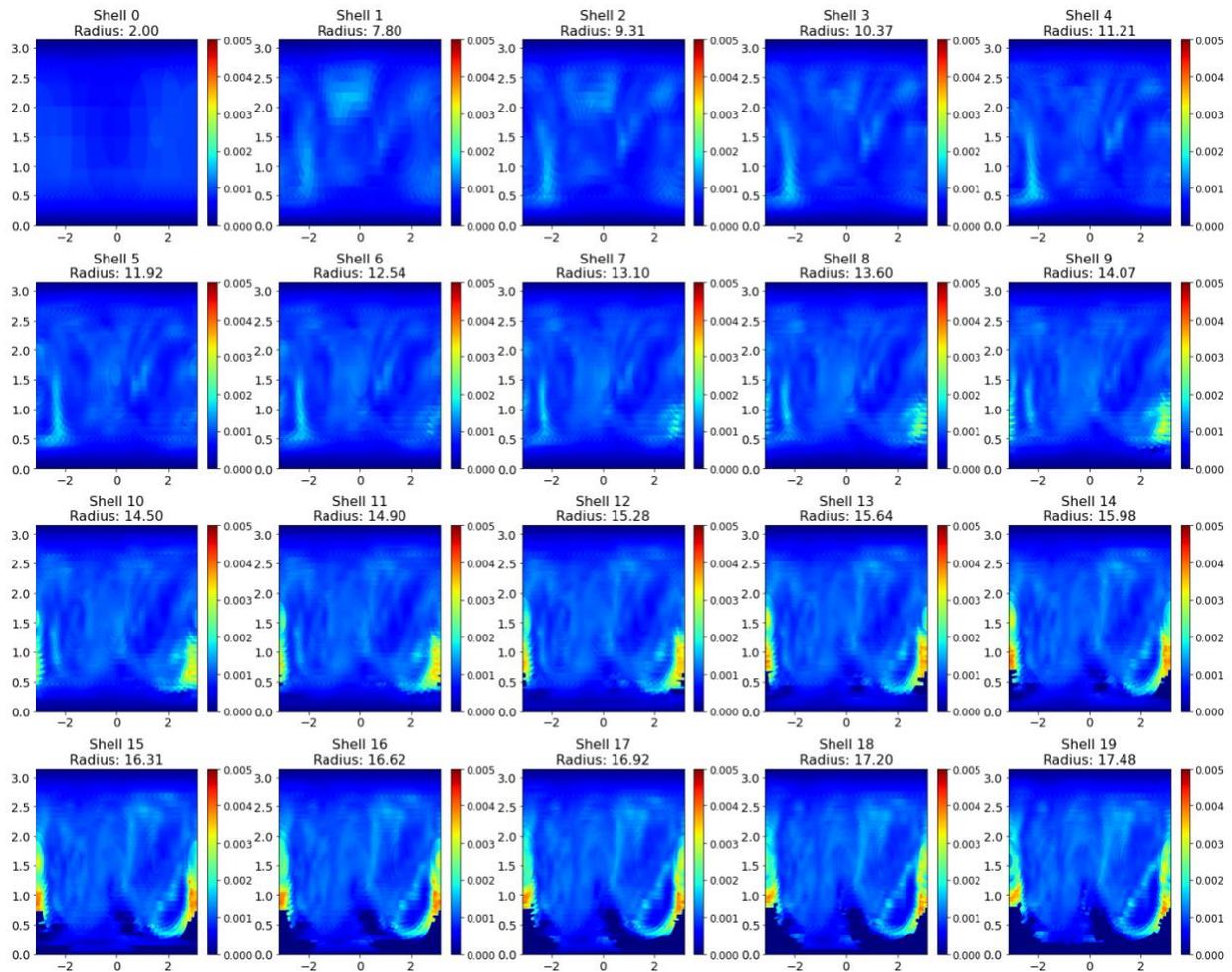

Figure C.3: Shell contour mapping for ADC in necrotic region

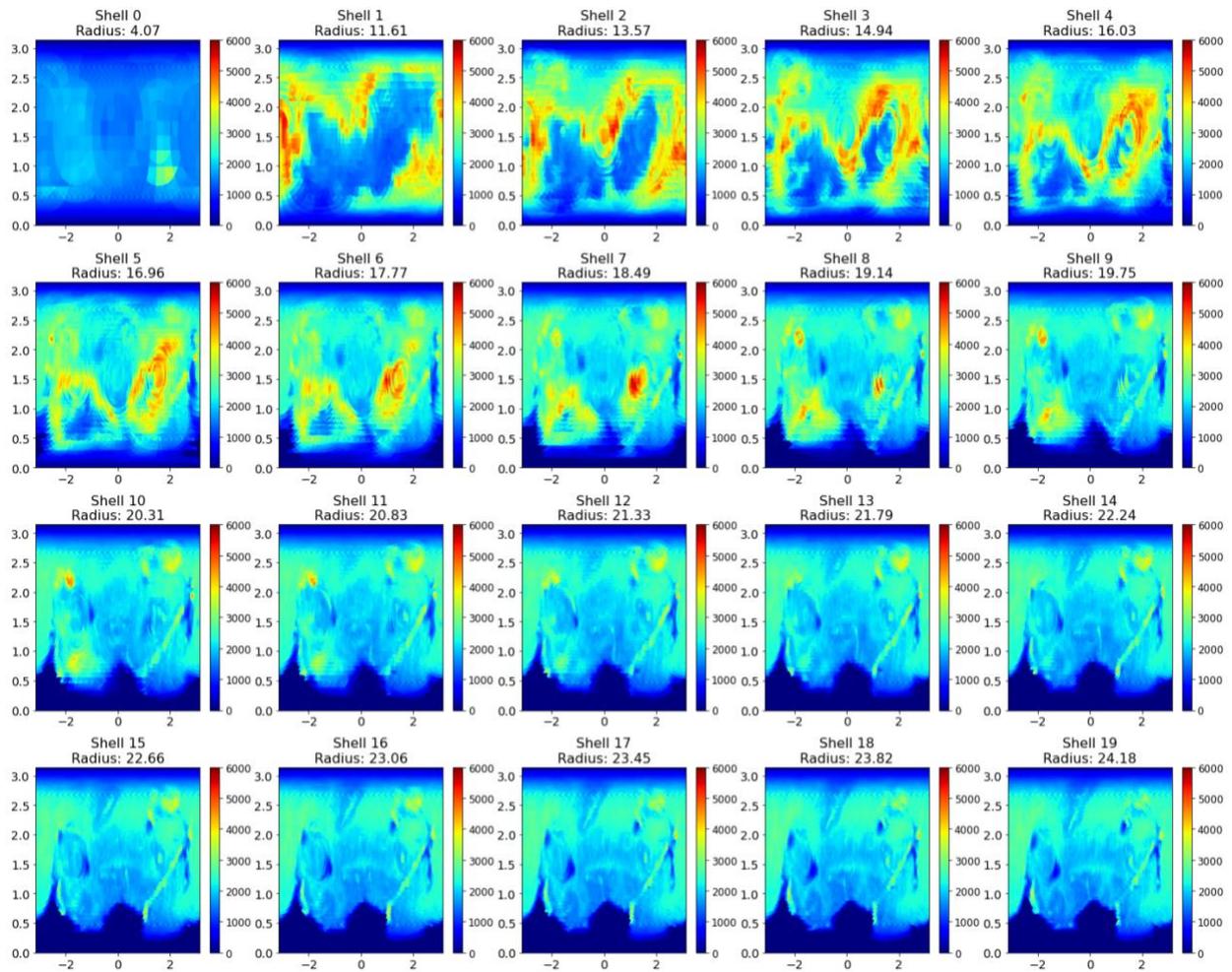

Figure C.4: Shell contour mapping for T1CE in T1 enhancing region

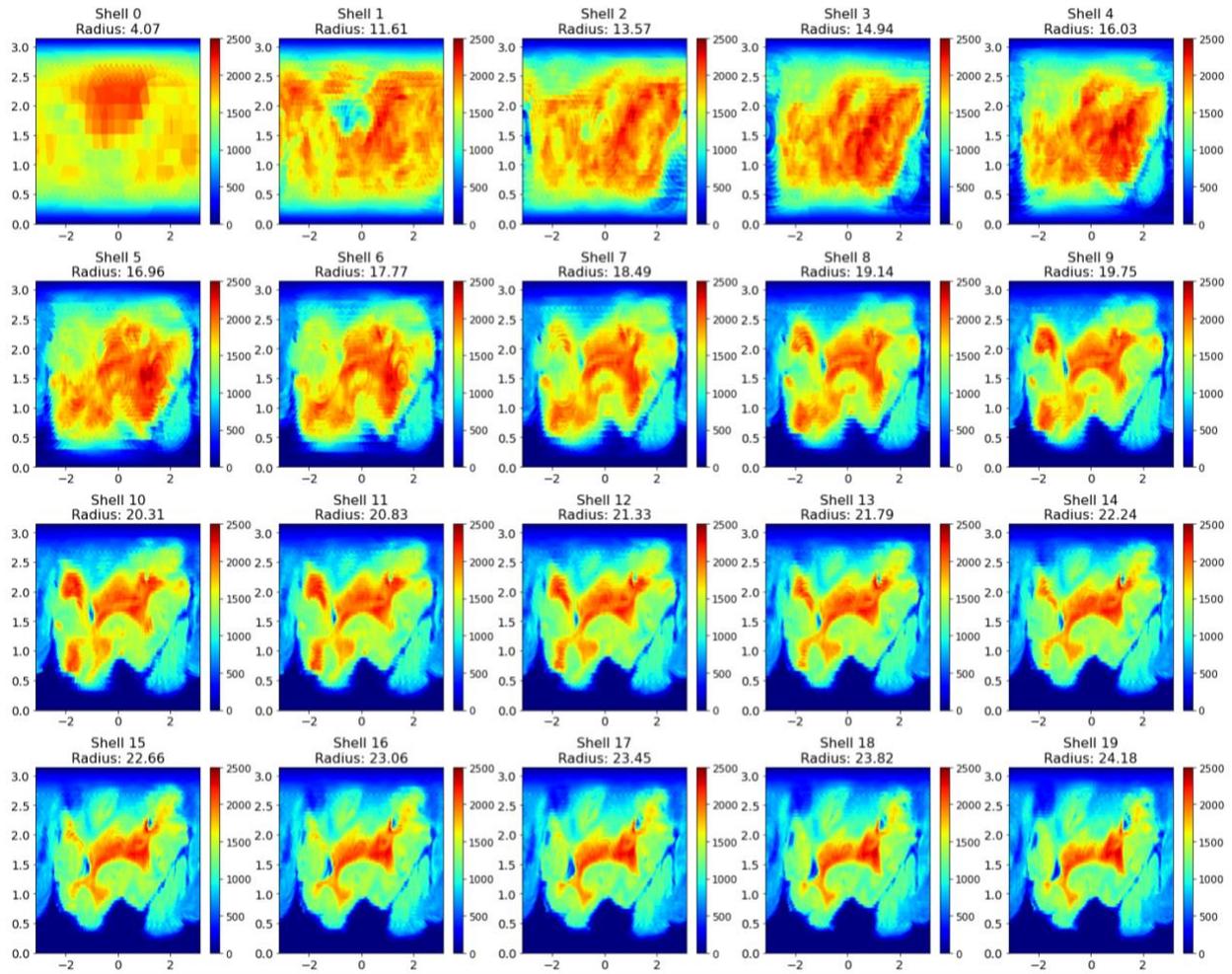

Figure C.5: Shell contour mapping for FLAIR in T1 enhancing region

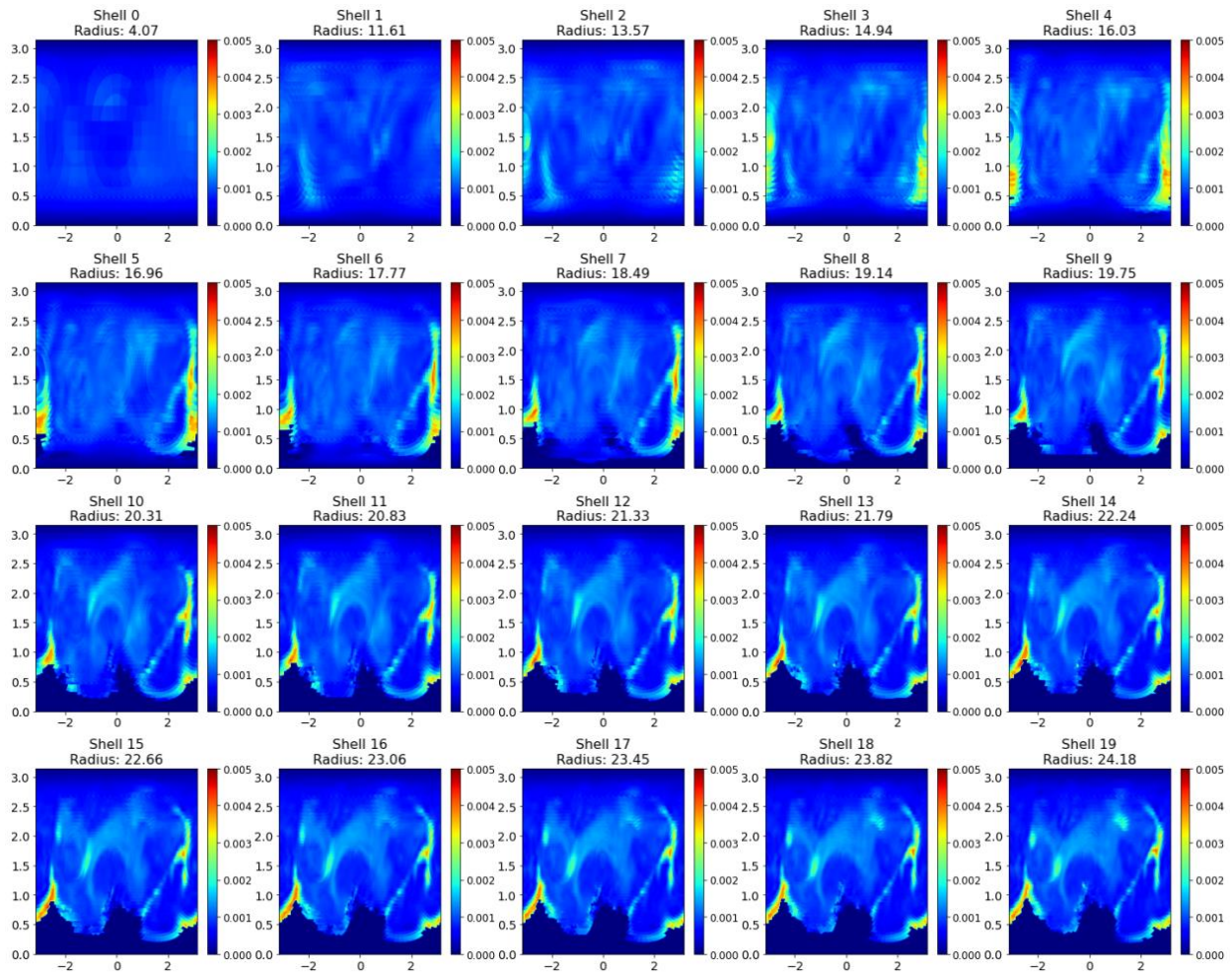

Figure C.6: Shell contour mapping for ADC in T1 enhancing region

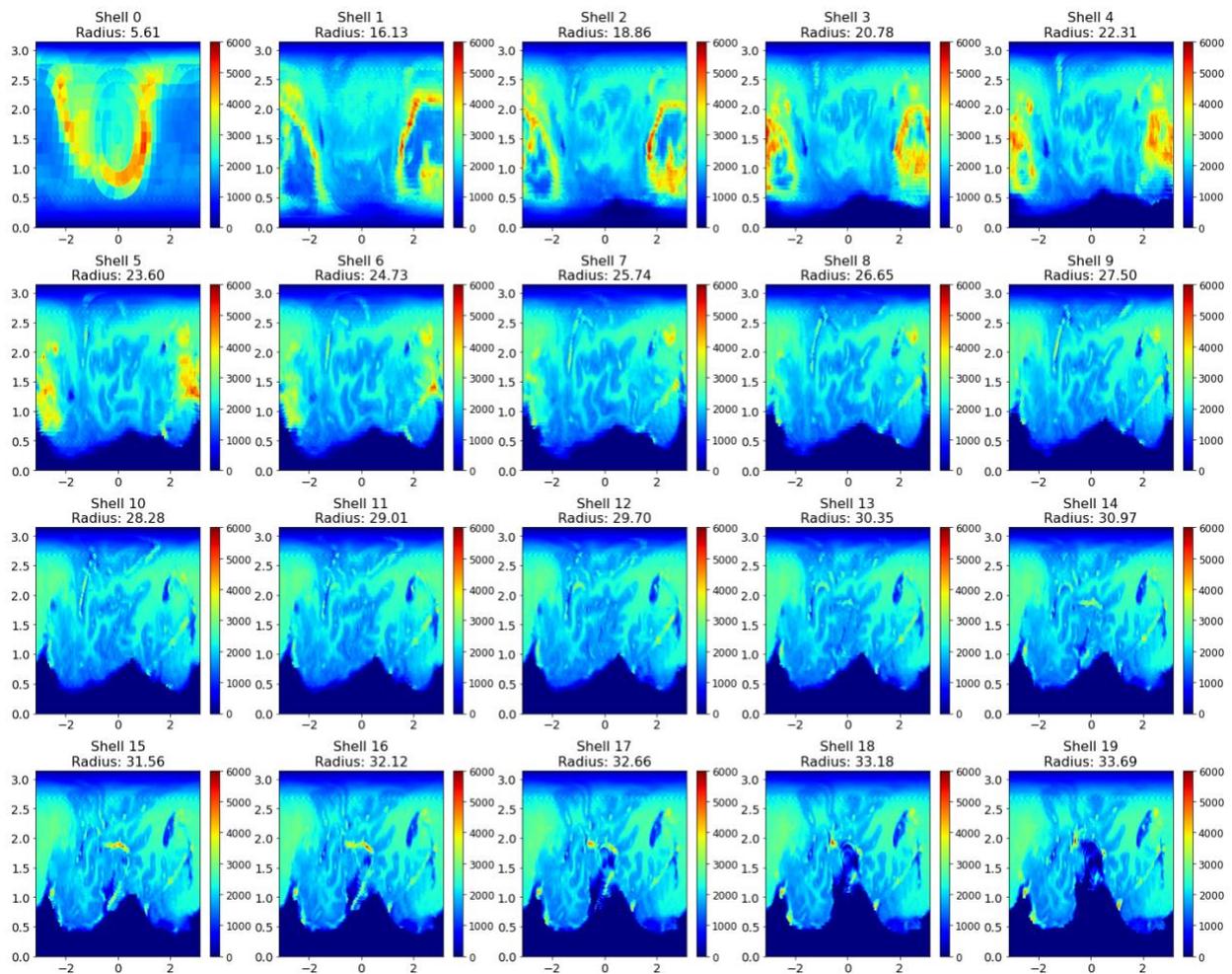

Figure C.7: Shell contour mapping for T1CE in T2 lesion region

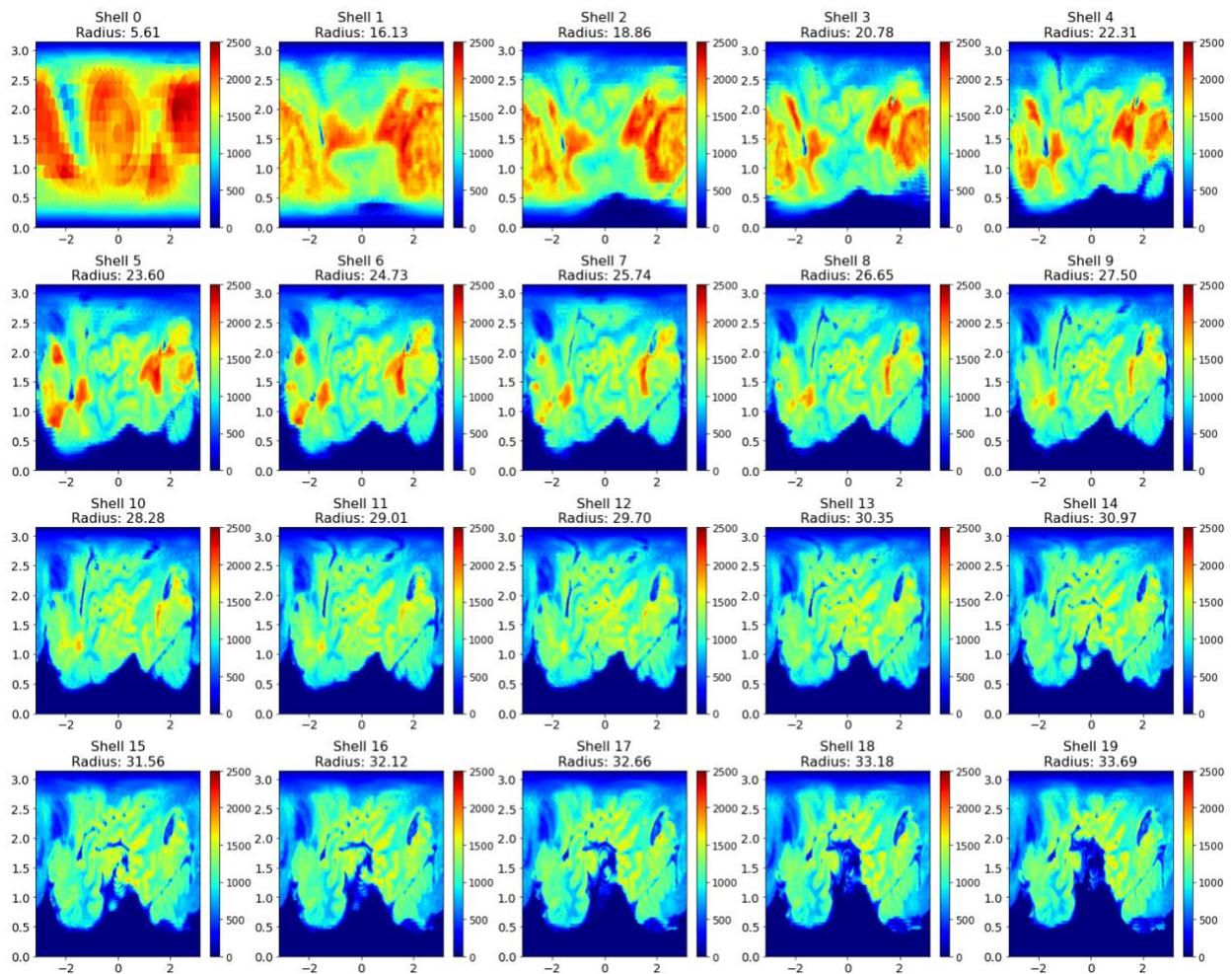

Figure C.8: Shell contour mapping for FLAIR in T2 lesion region

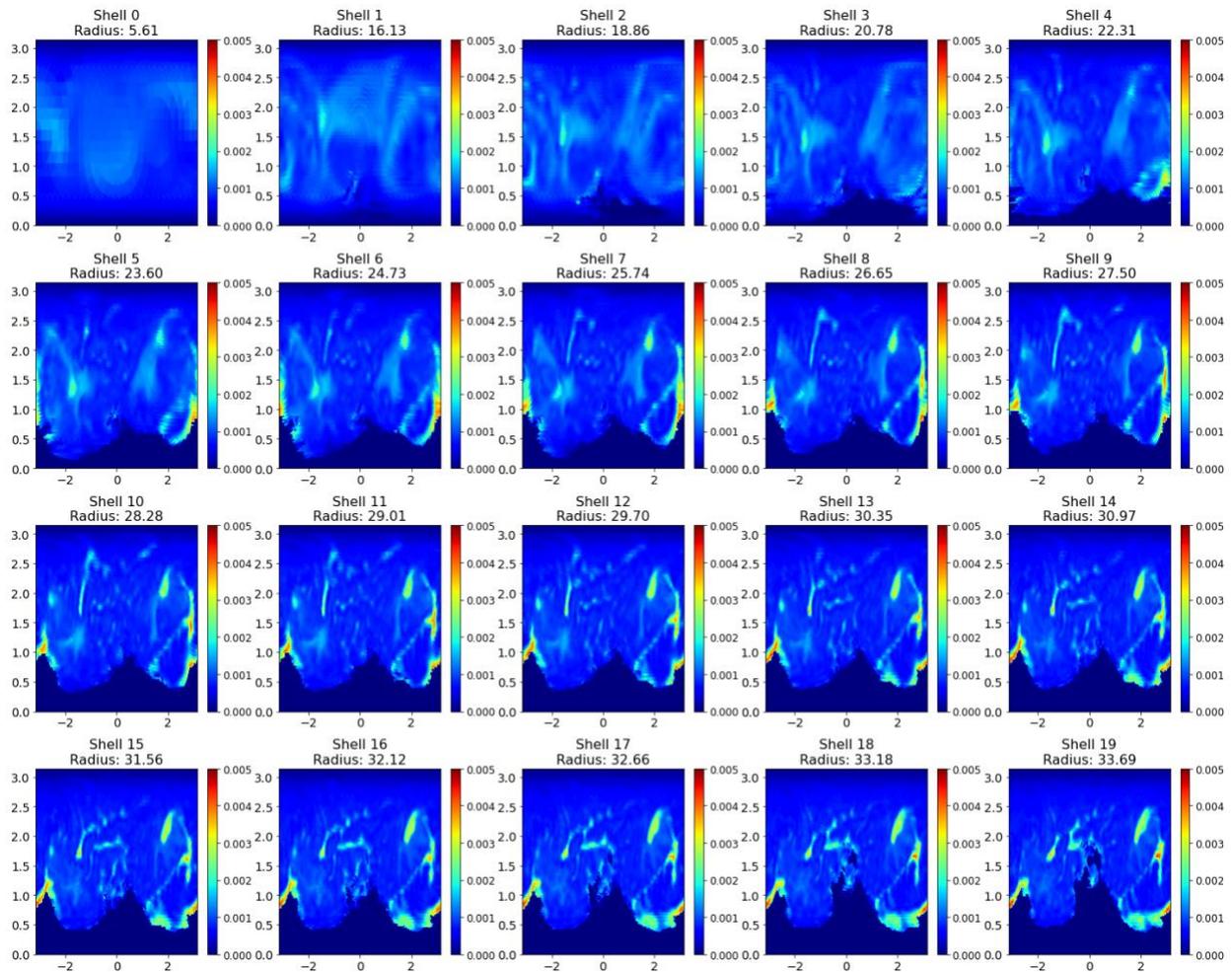

Figure C.9: Shell contour mapping for ADC in T2 lesion region